\documentclass[11pt]{article}
\usepackage{jheppub,amsmath, amsthm, amssymb,slashed,url,bbold,xcolor}
\usepackage{graphicx}
\usepackage{epstopdf}

\usepackage{etoolbox}% http://ctan.org/pkg/etoolbox
    \makeatletter
    \patchcmd{\maketitle}{\@fpheader}{}{}{}
    \makeatother

\def\nab{\overrightarrow{\nabla}}
\def\galnab{{\bf{\nabla}}}

\def\OMIT#1{{}}

\newcommand{\beq}{\begin{equation}}
\newcommand{\eeq}{\end{equation}}
\newcommand{\beqa}{\begin{eqnarray}}
\newcommand{\eeqa}{\end{eqnarray}}

\newcommand{\kf}{k_{\rm F}}
\def\kt{\widetilde k}

\begin{document}

%\preprint{\vbox{
%\psfig{file=mondrian_logo.ps,width=1.2in,angle=0}\hfill
%\psfig{file=hpc.ps,width=1.6in,angle=0}\hfill
%\psfig{file=JLab_logo.ps,width=2.1in,angle=0}\hfill
%\hbox{UNH-06-}}}
\dedicated{NT@UW-22-19}
%\phantom{ijk}
%\vskip 0.5cm
\title{Toward precision Fermi liquid theory in two dimensions}
\vskip 0.5cm
\author[a]{Silas R.~Beane,}
\author[b]{Gianluca Bertaina,}
\author[a]{Roland C.~Farrell}
\author[a]{and William R.~Marshall}

\affiliation[a]{Department of Physics, University of Washington,
  Seattle, WA 98195}
\affiliation[b]{Istituto Nazionale di Ricerca Metrologica, Strada delle Cacce 91, I-10135 Torino, Italy}

\vphantom{} \vskip 1.4cm \abstract{The ultra-cold and weakly-coupled
  Fermi gas in two spatial dimensions is studied in an effective field
  theory framework.  It has long been observed that universal corrections to the energy density to
two orders in the interaction strength do not agree with Monte Carlo
simulations in the weak-coupling regime. Here, universal corrections to three orders in the interaction strength
are obtained for the first time, and are shown to provide agreement between theory and
simulation. Special
  consideration is given to the scale ambiguity associated with the
  non-trivial renormalization of the singular contact interactions.
  The isotropic superfluid gap is obtained to next-to-leading order,
  and nonuniversal contributions to the energy density due to
  effective range effects, p-wave interactions and three-body forces
  are computed. Results are compared with precise Monte Carlo
  simulations of the energy density and the contact in the
  weakly-coupled attractive and repulsive Fermi liquid regimes. In
  addition, the known all-orders sum of ladder and ring diagrams
  is compared with Monte Carlo simulations at weak coupling and
  beyond.}\notoc

\maketitle
\newpage

%%%%%%%%%%%%%%%%%%%%%%%%%%%%%%%%%%%%%%%%%%%%%%%%%%%%%%%%%%%%%%%%%%%%%%%%%%%%%%%%%%%%%%%%
\section{Introduction}
\setcounter{page}{1}

\noindent Dramatic improvements in the experimental control of atomic
systems have led to intense theoretical interest in the quantum
mechanics of interacting atomic
gases.  In particular, the
ability to tune interaction strengths using Feshbach resonances, and
to continuously vary the number of spatial dimensions using
anisotropic harmonic traps and optical lattices, are allowing for
precision experimental tests of a vast and quickly-developing theoretical
framework.  This has attracted physicists from many areas of research
who are interested in the few- and many-body quantum mechanics of
non-relativistic constituents.  Simultaneous progress in numerical
simulation~\cite{Carlson:2012mh}, coupled with increased access to
high-performance computing, has been occurring in parallel with the
experimental developments.  This rich interplay among theory, experiment
and computation has led to what might be considered a golden age of
atomic physics.

An increasingly valuable tool for atomic systems that enables
model-independent descriptions of both bosonic and fermionic gases is
effective field theory (EFT)\footnote{For general reviews, see
  Refs.~\cite{Phillips:2002da,Kaplan:2005es}, and for an atomic-physics
  oriented review, see Ref.~\cite{Braaten:2004rn}.}.  Historically in
atomic physics, studies of gases with constituents interacting via
finite-range potentials have utilized specific solvable models of the
two-body interaction, like the hard-sphere potential. While these
models capture the essential physics of finite-range potentials, EFT
allows for the study of interacting gases in a manner that is
independent of any specific potential. The resulting interaction,
viewed in coordinate space, is a sequence of potentials consisting of
delta functions and their derivatives, which are highly singular near
the origin. However, the divergent nature of the interaction is
straightforward to control using regularization and renormalization,
and can be exploited by considering the renormalization group (RG)
flow of coupling constants. The main utility of the EFT framework is
that it provides a clear strategy for the systematic improvement of
the quantum mechanical descriptions of fundamental properties of
atomic gases at weak coupling. These improvements include, for
example, pairing and finite-temperature effects, nonuniversal
modifications to the equation of state, non-isotropic interactions,
many-body forces and shape-parameter corrections, as well as
dimensional-crossover scenarios. The EFT treatment of weakly-coupled
Fermi gases in three spatial dimensions has been developed in
Refs.~\cite{Furnstahl:1999ix,Steele:2000qt,Hammer:2000xg,Furnstahl:2002gt,Platter:2002yr,Schafer:2005kg,Schafer:2006yf,Furnstahl:2006pa,Kaiser:2011cg,Kaiser:2012sr,Chafin:2013zca,Kaiser:2013bua,Kaiser:2014laa}.
In this paper, these EFT techniques will be adapted and applied to the
study of weakly-coupled Fermi gases in two dimensions\footnote{For a
  comprehensive review of the Fermi gas in two dimensions, see
  Ref.~\cite{2015arca.book....1L}.}.

Experimentally, the Fermi gas in quasi-two dimensions is
accessible via highly-anisotropic harmonic traps that effectively
confine a spatial dimension. This requires {\it inter alia} a special
theoretical treatment which accounts for the continuous
compactification of the third dimension. However, recent numerical
simulations using Monte Carlo (MC) techniques allow a precise, direct
computation of the zero-temperature equation of state in two spatial
dimensions from weak coupling all the way to the BCS-BEC crossover
region~\cite{PhysRevLett.106.110403,BertainaS,PhysRevA.92.033603,Galea:2015vdy,Galea:2017jhe,Rammelmuller:2015ybu}. 
It has been observed that the energy density obtained from these simulations is in tension with theoretical calculations at weak coupling, 
which have been carried out to second order in the universal coupling~\cite{PhysRevB.45.10135,PhysRevB.45.12419,PhysRevB.12.125}.
In this paper, the calculation of the energy density is carried out to third order in the universal coupling and is found to resolve the tension between theory and simulation.

Consider a gas of fermions of mass $M$ in two or three spatial dimensions with two-body
interactions that are of finite range $R$ and typical energy scale $U$. The dimensionless parameters
$\kf R$ and $M U/\kf^2$, where $\kf$ is the Fermi momentum, are the knobs which
determine the strength of the interaction.  Without fine tuning one expects that the ground state of the Fermi gas
will be dominated by the leading s-wave two-body interactions, which are governed by the two- (three-)dimensional scattering lengths, $a_2$ ($a_3$).
Unlike the case with $a_3$, whose sign in the weakly-interacting regime is indicative of whether the interaction is attractive or repulsive, $a_2$ is intrinsically positive.
The relevant dimensionless
parameter is $\kf a_2$ ($\kf a_3$) and
weak
coupling is therefore achieved either at low density (dilute limit),
or with weak two-body interactions. Due to the presence of a Fermi
surface, an arbitrarily weak attractive interaction leads to the
formation of Cooper pairs and qualitatively changes the properties of
the gas.  At weak coupling, the gas is in the BCS phase characterized
by large interparticle spacing $\kf^{-1}$ and exhibits superfluidity
or superconductivity. As the coupling increases, there is a transition
to the BEC phase of tightly bound pairs, and it becomes natural to
view the gas as a system of weakly-repulsive bosonic dimers.  Here
dimensionality provides a drastic
difference~\cite{2015arca.book....1L}.  Whereas in three dimensions,
attraction must be strong enough to sustain a bound state, in two
dimensions there is a bound state for an arbitrarily weak interaction.
This implies that in two dimensions the entire BCS-BEC crossover may
be traversed by varying the density with an arbitrary attractive
interaction i.e. there are no new singularities introduced due to the
formation of a bound state. 

Many of the interesting and distinguishing
features of the two-dimensional gas arise from the manner in which
quantum effects break the scale invariance of the classical
Hamiltonian and give rise to the effective coupling constant
$-1/\log\left(\kf a_2\right)$. This paper will be concerned with
the weak-coupling limits: $|\log\left(\kf a_2\right)|\gg 1$.  With
repulsive interactions the gas is a Fermi liquid with the energy
density a straightforward perturbative expansion in the coupling. With
attractive interactions the gas is a paired superfluid, but may be
viewed as a Fermi liquid as long as the energy due to pairing is small
as compared to the leading perturbative correction to the free Fermi
gas. The various scale hierarchies relevant for a complete and
systematic description of the weak-coupling regime may be precisely
quantified in the EFT.

Beyond mean-field corrections to the energy of the weakly-coupled
Fermi gas in two dimensions with repulsive interactions were first
considered in
Refs.~\cite{PhysRevB.12.125,PhysRevB.45.10135,PhysRevB.45.12419}.  The
attractive Fermi gas and the superfluid gap were treated in
Ref.~\cite{PhysRevLett.62.981,PhysRevB.41.327}.  The goal of this
paper is to consider the next order in the weak-coupling expansion,
and to compare the results with MC simulations. It is
important to stress that the subleading corrections that we compute
have previously been found in two distinct studies whose aim was to
perform resummations of classes of Feynman\footnote{Here all diagrams,
  both in free space and in medium, are referred to as Feynman
  diagrams.}  diagrams to all orders in the
coupling~\cite{Kaiser:2013bua,Kaiser:2014laa}. A secondary goal of
this paper is to obtain the leading nonuniversal effects, due to
effective range corrections, p-wave interactions and three-body
forces, although it is not clear whether these latter two effects can be
meaningfully compared with experiment or simulation at the present
time. Indeed, the motivation for computing these effects is to inspire
MC simulations which include more intricate few-body forces and enable
a meaningful comparison with theory.

This paper is organized as follows.  Sec.~\ref{sec:EFTtech} introduces
the effective Lagrangian density which encodes the interactions that
are the basis of the necessary EFT technology, and considers the
free-space power-counting scheme.  In Sec.~\ref{sec:TBFS}, a general
partial-wave expansion of the two-body scattering amplitude is
given. With the assumption of finite-range forces, effective range
expansions of the s- and p-wave scattering amplitudes are
defined. Finally, the scattering amplitudes are reproduced in the EFT
using dimensional regularization.  The modifications of the free-space
EFT technology to the treatment of interactions in medium are
considered in Sec.~\ref{sec:FDT}.  This section adapts the main
results of Refs.~\cite{Hammer:2000xg,Furnstahl:2006pa} to the case of
two spatial dimensions. In particular, the renormalized thermodynamic
potential is obtained, and the superfluid gap is recovered
using dimensional regularization. The Fermi liquid expansion is
treated to three orders in the expansion parameter in
Sec.~\ref{sec:DFGu}.  Nonuniversal corrections to the energy density
are considered in Sec.~\ref{sec:DFGnu}.  In Sec.~\ref{sec:summandgen},
the final expression of the energy density is given at an arbitrary
renormalization scale, and the contact is defined and obtained from the
energy density. In Sec.~\ref{sec:LaR}, resummation schemes which treat
ladder~\cite{Kaiser:2013bua} and ring~\cite{Kaiser:2014laa} diagrams
to all orders in the interaction strength are reviewed. Comparison of
theoretical predictions of the weak-coupling regime with MC simulations is
given in Sec.~\ref{sec:MCS}.  Sec.~\ref{sec:conc} is a summary and
conclusion.

\section{Effective field theory technology}
\label{sec:EFTtech}

\subsection{Effective Lagrangian}
\label{sec:EL}

\noindent Here it is assumed that the underlying interaction
experienced by the fermionic atoms is of finite range, say $R$. Then,
with a characteristic momentum scale represented by $k$, at momentum
scales $k\leq R^{-1}$, the interaction takes the form of a sequence of
contact interactions. The theory of contact interactions is described
by an effective Lagrangian which consists of local operators
constructed from the non-relativistic fermion field $\psi$, which
generally possesses $g$ components\footnote{Note that the three-dimensional relationship between degeneracy and spin, $g=2s+1$, holds
when two dimensions is reached as a limiting case via dimensional reduction.}. The local Lagrangian density is constrained to be
Galilean and time-reversal invariant and can be expressed in the form
\beqa
  {\cal L}  &=&
       \psi^\dagger \bigg\lbrack i\partial_t + \frac{\nab^{\,2}}{2M}\bigg\rbrack
                 \psi - \frac{C_0}{2}(\psi^\dagger \psi)^2
            + \frac{C_2}{16}\Big\lbrack (\psi\psi)^\dagger 
                                  (\psi\galnab^2\psi)+\mbox{ h.c.} 
                             \Big\rbrack
  \nonumber \\[5pt]
   & & \null + \:
         \frac{C_2'}{8} (\psi \galnab_i \psi)^\dagger 
             (\psi\galnab_i \psi) 
            - \frac{C_4'}{64}\Big\lbrack (\psi \nabla_i \psi)^\dagger 
                                  (\psi\galnab^2 \nabla_i \psi)+\mbox{ h.c.} 
                             \Big\rbrack
         - \frac{D_0}{6}(\psi^\dagger \psi)^3 +  \ldots\ ,
  \label{lag}
\eeqa
where $\galnab=\overleftarrow{\nabla}-\nab$ is the Galilean invariant
derivative and h.c.\ symbolizes the hermitian conjugate.  Throughout
the paper, $\hbar =1$, and the fermion mass, $M$, is left explicit.
The Lagrangian takes the same form in any spacetime dimension $d$,
with the dimensions of the fermion field and of the
operator coefficients given by $[\psi]=(d-1)/2$, $[C^{(\prime)}_{2n}]=2-d-2n$,
and $[D_{2n}]=3-2d-2n$.

In two spatial dimensions ($d=2+1$), the Lagrangian with only the $C_0$
interaction is scale invariant\footnote{In fact, this theory is also
  invariant under non-relativistic conformal transformations, see
  Ref.~\cite{Jackiw:1991je}.}.  This is most easily seen by rescaling
the field and spatial coordinates by $\psi \rightarrow M^{1/2}\psi$
and ${\bf x}\rightarrow M^{-1/2} {\bf x}$, which removes all
dimensionful parameters from the Hamiltonian obtained from
Eq.~(\ref{lag}). The resulting theory has a marginal contact
interaction whose strength is proportional to the dimensionless
coupling $ M C_0$. This symmetry does not survive quantization and is
broken by the regularization of the singular $C_0$ interaction which
necessarily introduces a scale into the theory.  This constitutes a
fundamental difference between the many-body physics of two and three
dimensions.

\subsection{Free-space counting scheme}
\label{sec:EFTcount}

\begin{figure}[!h]
\centering
\includegraphics[width = 0.5\textwidth]{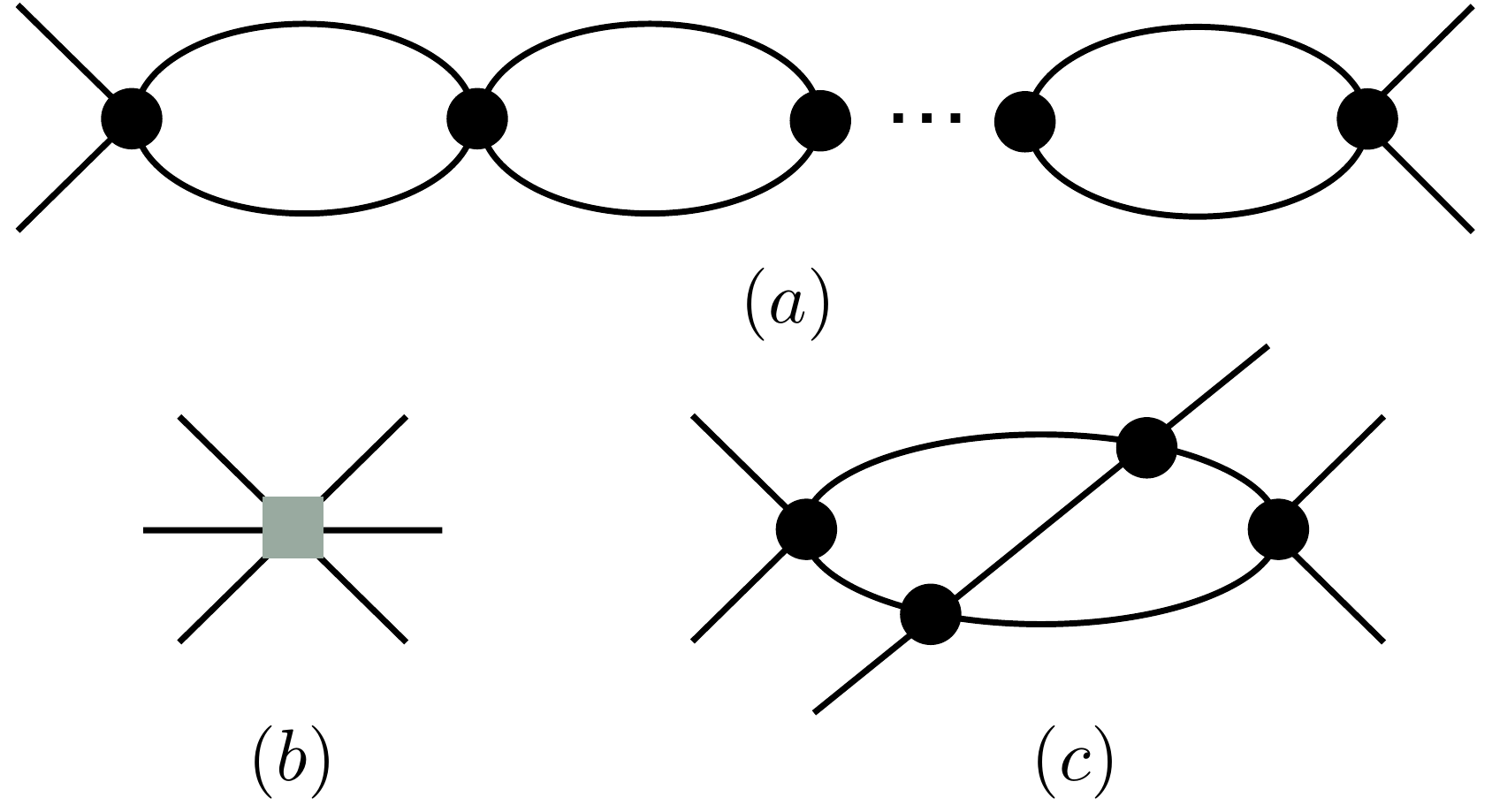}
\caption{Two-body diagram with $L$ loops (a). Three-body diagrams at tree level (b) and at two-loop level (c) (fish slash). The black circle (grey square) corresponds
  to an insertion of the $C_0$ ($D_0$) operator.}
    \label{fig:PCfree}
\end{figure}

\noindent By exploiting topological properties, it is found that a free-space Feynman diagram
with $L$ loops or $E$ external lines and $V_{2i}^n$ $n$-body vertices with $2i$ derivatives scales as
$(k R)^\chi$, where 
\begin{eqnarray}
\chi & = & (d-1)L+2+\sum_{n=2}^\infty \sum_{i=0}^\infty (2i-2) V_{2i}^n \nonumber \\
   &= & d+1 - \frac{1}{2}(d-1)E + \sum_{n=2}^\infty \sum_{i=0}^\infty \left(2i+(d-1)n-d-1\right) V_{2i}^n \ .
        \label{eq:kpowercount}
\end{eqnarray}
The EFT has predictive power in three dimensions because at every
order in $k R$ there are a finite number of diagrams that
contribute. In two dimensions there is a subtlety due to the scale
invariance of the universal interaction (i.e. the effect of the $C_0$
operator), as noted above. In order to distinguish two dimensions from
three, it is instructive to power count the generic interactions
illustrated in Fig.~\ref{fig:PCfree}. The two-body diagram with $L$
loops and $L+1$ insertions of $C_0$, Fig.~\ref{fig:PCfree}(a), has
$\chi=L(d-3)$. Therefore, in three dimensions, there is a loop
expansion\footnote{This assumes that the $C_0$ operator is of natural
  size. Near unitarity, each $C_0$ insertion brings an infrared
  enhancement of $k^{-1}$, leading to all loops counting equally and a
  consequent breakdown of perturbation
  theory~\cite{vanKolck:1998bw,Kaplan:1998tg,Kaplan:1998we}.} with
each loop bringing one additional power of $k R$. By contrast, in two
dimensions, the two-body diagram with $L$ loops has $\chi=0$, a
consequence of scale invariance, and an indication that universal
interactions appear in a perturbative expansion in $M C_0$.  A further
illustrative example is the leading three-particle interactions.  The
three-body diagram with an insertion of $D_0$,
Fig.~\ref{fig:PCfree}(b), has $\chi=0$ for all $d$. A leading
three-body diagram with four $C_0$ insertions,
Fig.~\ref{fig:PCfree}(c), has $\chi=2(d-4)$.  Therefore, in three
dimensions these three-body effects appear at the same order in the
momentum expansion, as is necessary given that the two-loop diagram
has a logarithmic divergence, which must be renormalized by the $D_0$
operator~\cite{PhysRevB.55.8090}. By contrast, in two spatial
dimensions, these universal three-body diagrams require no new
counterterms beyond $C_0$, and indeed they appear at lower order in
the power counting, indicating that three-body forces are enhanced in
two dimensions.

\section{Two-bodies in free space}
\label{sec:TBFS}

\subsection{Partial-wave expansion}
\label{sec:PWE}

\noindent Consider two-body scattering, with incoming momenta labeled
${\bf k}_1,{\bf k}_2$ and outgoing momenta labeled ${\bf k}'_1,{\bf
  k}'_2$. In the center-of-mass frame, ${\bf k}\equiv{\bf k}_1=-{\bf k}_2$, ${\bf k}'\equiv{\bf k}_1'=-{\bf k}_2'$
and $k\equiv|{\bf k}| = |{\bf k}'|$.  In two dimensions, angular momentum is specified
by counting the number of windings around the unit circle,
including both clockwise and anti-clockwise orientations. The unitary
scattering amplitude can be expanded in partial waves
as~\cite{adhikari,Rakityansky_2012,Hammer:2010fw}
\begin{eqnarray}
  T(k,\phi) & = & \sum_{\ell=0}^\infty T_\ell(k,\phi) \ =\ \frac{4}{M} \sum_{\ell=0}^\infty \frac{\epsilon_\ell \cos\left( \ell\phi \right)}{\cot\delta_\ell(k)-i}    \ ,
  \label{eq:PW2dexp2}
\end{eqnarray}
where $\phi$ is the scattering angle, $\delta_{\ell}$ is the phase shift,
$\epsilon_0=1$ and $\epsilon_\ell=2$ for $\ell>0$ and the normalization has been chosen to 
match the Feynman diagram expansion.

The $\ell=0$ (s-wave) scattering amplitude is then
\begin{eqnarray}
  T_0(k) & = &  \frac{4}{M}\frac{1}{\cot\delta_0(k)-i} \ .
  \label{eq:swaveSA}
\end{eqnarray}
The effective range expansion takes the conventional form
\begin{eqnarray}
\cot\delta_0(k) \ = \ \frac{1}{\pi} \log{\left({k^2}{a_2^2}\right)} \ +\ \sigma_2\, k^2 \ + \ {\it O}( k^4 ) \ ,
\label{eq:9mod}
\end{eqnarray}
where $a_2$ is the scattering length\footnote{Another common
  convention for the definition of the scattering length coincides with the diameter $a_{2D}$ in the case of
 the hard-disk potential, corresponding to $a_{2D}=2a_2\exp{(-\gamma)}$, where $\gamma$ is the Euler-Mascheroni constant.}
and $\sqrt{|\sigma_2|}$ is the effective range. This form of the expansion appears
odd from the EFT perspective since the leading effect at low-$k$ is non-analytic
in $k$. As will be seen below, this purely quantum mechanical effect occurs because
of strong infrared effects in two dimensions.

The $\ell=1$ (p-wave) scattering amplitude is
\begin{eqnarray}
  T_1(k,\phi) & = & {\bf k}\cdot{\bf k}' \frac{8}{M}\frac{1}{k^2\cot\delta_1(k)-i k^2} \ ,
  \label{eq:pwaveSA}
\end{eqnarray}
and the p-wave effective range expansion can be written as
\begin{eqnarray}
  k^2\cot\delta_1(k) &=& -\frac{1}{\sigma_p} \ +\ \frac{1}{\pi}k^2\log{\left({k^2}{a_p^2}\right)} \ + \ {\it O}( k^4 ) \ ,
   \label{eq:pwaveERE}
\end{eqnarray}
where $\sigma_p$ is a scattering volume (units of area), and $a_p$ is a length scale that characterizes higher order effects in the momentum expansion.
For $\sigma_p k^2\ll 1$, the scattering amplitude can be expanded in perturbation theory to give
\begin{eqnarray}
  T_1(k,\phi) & = & -\sigma_p {\bf k}\cdot{\bf k}' \frac{8}{M}\bigg\lbrack  1 \ +\  \frac{1}{\pi}\sigma_p k^2\Big( \log{\left({k^2}{a_p^2}\right)} \ -\ i {\pi} \Big)
  +{\it O}\left((\sigma_p k^2)^2\right)\bigg\rbrack \ .
  \label{eq:pwaveSApt}
\end{eqnarray}
In the next subsection, these scattering amplitudes will be recovered in the EFT.

\subsection{Scattering in the EFT: s-wave}
\label{sec:IS}

\begin{figure}[!h]
\centering
\includegraphics[width = 0.5\textwidth]{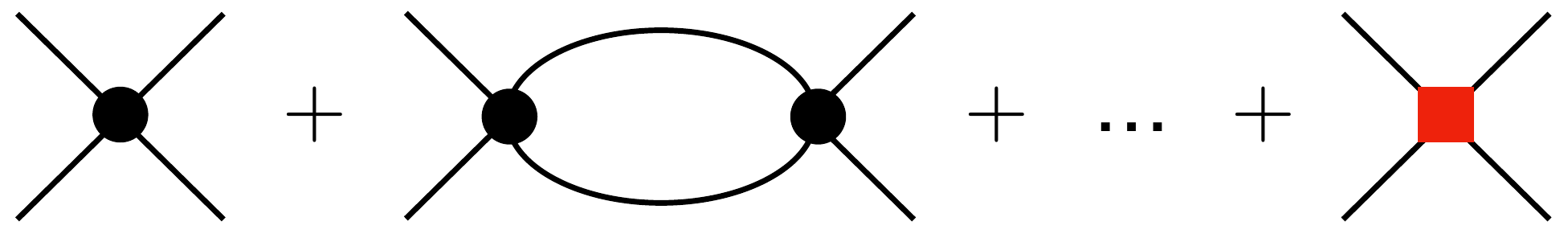}
\caption{Diagrams contributing to isotropic scattering. The black circle (red square) corresponds to an insertion of the $C_0$ ($C_2$) operator.}
    \label{fig:scatt}
\end{figure}

\noindent Consider s-wave scattering in the EFT described by the
effective Lagrangian of Eq.~(\ref{lag})\footnote{This section
  closely follows the development in
  Refs.~\cite{Beane:2010ny,Beane:2018jyq}.}. The sum of the Feynman
diagrams shown in Fig.~\ref{fig:scatt} is
\begin{equation}
T_0(k) \ = \ - C_{0}  \ - \ C^2_{0}\,I_0(k)   \ +\ \ldots \ - \ C_2 k^2  \ ,
\label{eq:T1}
\end{equation}
where
\begin{eqnarray}
I_0(k) \ = \ M\left(\frac{\mu}{2}\right)^{\epsilon} \int \! \frac{d^{d-1}{\bf q}}{
  (2\pi)^{d-1}}
\frac{1}{ k^2-{{q}^2} + i \delta}
\ ,
\label{eq:2b}
\end{eqnarray}
$\mu$ is the dimensional regularization (DR) scale, and $\epsilon\equiv 3-d\,$\footnote{In DR the couplings are multiplied by $\left (\frac{\mu}{2}\right )^{\epsilon}$ to keep the action dimensionless. 
In the EFT, this is equivalent to multiplying the divergent loop integrals by $\left (\frac{\mu}{2}\right )^{\epsilon}$.}. As
perturbative physics can always be treated nonperturbatively, it is
convenient to neglect the $C_2$ contribution, and sum the bubble chain to all orders,
giving
\begin{eqnarray}
T_0(k) & = & -\frac{ C_{0}}{
  1 - I_0(k)  C_{0} } \ .
\label{eq:T2}
\end{eqnarray}
A useful integral is:
\begin{eqnarray}
\openup3\jot
I_n(k)&=& M\left({\mu\over 2}\right)^{\epsilon} \int \! {d^{d-1}{\bf  q}\over (2\pi)^{d-1}}\, 
{q}^{2n} \left({1\over k^2  -{q}^2 + i\delta}\right) 
\nonumber\\
&=&  k^{2n}\frac{M}{4\pi}\Bigg\lbrack \log{\left(-\frac{k^2}{\mu^2}\right)}\ +\ \gamma \ - \log\pi\ -\ \frac{2}{\epsilon} \Bigg\rbrack  \nonumber \\
&=& k^{2n} I_0(k) \ ,
\label{eq:2a}
\end{eqnarray}
where $n$ is a non-negative integer and the logarithmic divergence has been captured by the $1/\epsilon$ pole.

Using Eq.~(\ref{eq:T2}) and Eq.~(\ref{eq:2a}) gives
\begin{eqnarray}
  T_0^{-1}(k) & = & -\frac{1}{C_0}\ +\ 
\frac{M}{4\pi} \Bigg\lbrack \log{\left(-\frac{k^2}{\mu^2}\right)}\ +\ \gamma \ - \log\pi\ -\ \frac{2}{\epsilon} \Bigg\rbrack \ .
  \label{eq:T3}
\end{eqnarray}
Defining the renormalized EFT coefficient $C_{0}(\mu)$ with $\overline{MS}$ results in
\begin{eqnarray}
 -\frac{1}{C_0} & \equiv & -\frac{1}{C_0(\mu)}\ -\ 
\frac{M}{4\pi} \Bigg\lbrack \ \gamma \ - \log\pi\ -\ \frac{2}{\epsilon} \Bigg\rbrack \ .
  \label{eq:MSbar}
\end{eqnarray}
This exact renormalization condition then gives the physical scattering amplitude,
\begin{eqnarray}
  T_0^{-1}(k) & = & -\frac{1}{C_0(\mu)}\ +\ \frac{M}{4\pi} \log{\left(\frac{k^2}{\mu^2}\right)}\ -\ i \frac{M}{4} \ ,
  \label{eq:T3B}
\end{eqnarray}
with $C_{0}(\mu)$ treated to all orders. For the perturbative calculations presented below, it is useful
to expand Eq.~(\ref{eq:MSbar}) formally to third order in the renormalized coupling,
\begin{eqnarray}
C_0 & = & {C_0(\mu)}\Bigg\lbrace 1\ -\ 
\frac{M C_0(\mu)}{4\pi} \Bigg\lbrack \ \gamma \ - \log\pi\ -\ \frac{2}{\epsilon} \Bigg\rbrack \ +\ \nonumber \\
&&\qquad\qquad\qquad\left(\frac{M C_0(\mu)}{4\pi} \Bigg\lbrack \ \gamma \ - \log\pi\ -\ \frac{2}{\epsilon} \Bigg\rbrack\right)^2 \ +\ {\it O}( C_0(\mu)^3 )\Bigg\rbrace \ .
\label{eq:MSbarpert}
\end{eqnarray}

Now consider matching Eq.~(\ref{eq:T3B}) to the effective range expansion given in Eq.~(\ref{eq:9mod}).
In order to include effective range corrections via the $C_2$ operator, it is convenient to note that
in DR all contact interactions can be formally summed to give
\begin{eqnarray}
T_0(k) & = & -{ \sum C_{2n} \ k^{2n}  \over
1 - I_0(k) \sum C_{2n} \ k^{2n}} \ ,
\label{eq:2gen}
\end{eqnarray}
with
\beq
\cot\delta_0(k) \ = \ \frac{1}{{\rm Im} \,I_0(k) }\Bigg\lbrack\frac{1}{\sum C_{2n} \ k^{2n}}\ -\  {\rm Re} \,I_0(k) \Bigg\rbrack \ .
\label{eq:5}
\eeq
Matching Eq.~(\ref{eq:T3B}), with the $C_2$ contribution included, to the effective range expansion then gives the EFT-inspired form:
\beq
\cot\delta_0(k) \ = \frac{2}{\pi}\bigg\lbrack \log{\left(\frac{k}{\mu}\right)} \ -\ \frac{1}{\alpha(\mu)}\bigg\rbrack\ +\ \sigma_2\, k^2 \ + \ {\it O}( k^4 ) \ ,
\label{eq:match1}
\eeq
where
\begin{eqnarray}
\alpha(\mu )&\equiv& \ \frac{M C_0(\mu )}{2\pi} \ =\ -\frac{1}{\log \mu a_2}\ \ \ ,  \ \ \ \sigma_2\ = \ \frac{4C_2(\mu)}{M C_0(\mu )^2} \ .
\label{eq:9b}
\end{eqnarray}
As intuited above from general scaling arguments, it is clear that $\alpha(\mu)$ is a dimensionless scale-dependent coupling constant which is the natural expansion parameter in the EFT.
\begin{figure}[!h]
\centering
\includegraphics[width = 0.75\textwidth]{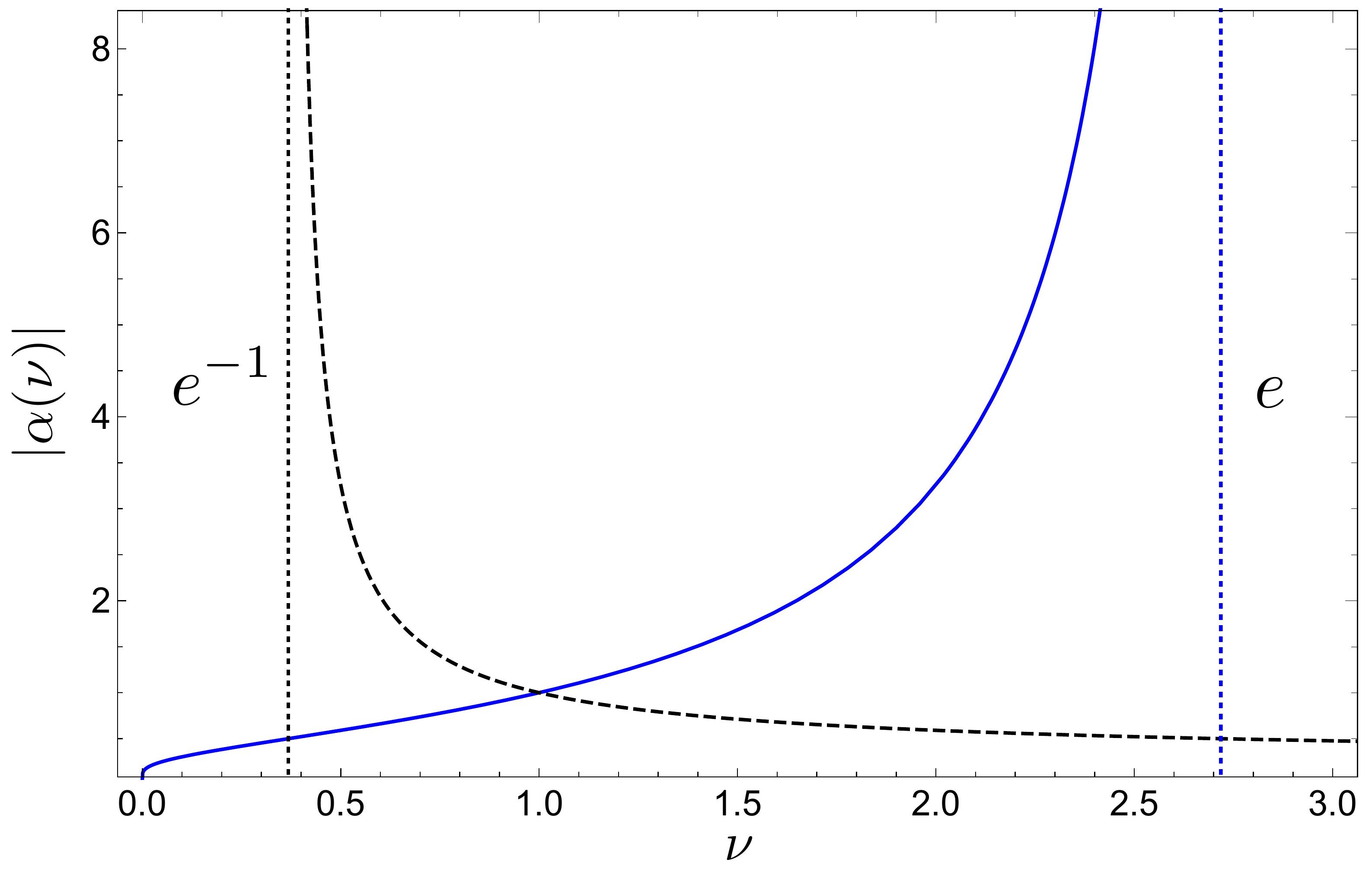}
\caption{Running of the coupling with $\mu=1$ and $|\alpha(1)|=1$, as described in the
  text: the solid blue curve corresponds to repulsive coupling,
  $\alpha(\nu)=|\alpha(\nu)|$, while the dashed black curve
  corresponds to attractive coupling,
  $\alpha(\nu)=-|\alpha(\nu)|$. The vertical dotted (blue) line to the
  right corresponds to the position of the Landau pole at $\nu=e$ in
  the repulsive case, while the vertical dotted (black) line to the
  left corresponds to the bound state at $\nu=e^{-1}$ in the
  attractive case.}
    \label{fig:alpharun}
\end{figure}

Evidently the condition for a bound state, $\cot\delta_0(i \gamma_B)=i$ with binding momentum $\gamma_B>0$, is satisfied both for attractive coupling, $C_0<0$, and for repulsive coupling, $C_0>0$. This is
due to the strong infrared quantum effects which give rise to the logarithm when $C_0$ is treated to all orders.
Neglecting range corrections, the binding momentum is given by:
\begin{eqnarray}
\gamma_B &=& \mu \exp\left(\frac{1}{\alpha(\mu)}\right) \ =\ 1/a_2 \ ,
\label{eq:bmswave}
\end{eqnarray}
with binding energy $\varepsilon_B=-\gamma_B^2/M$.

The RG evolution of $\alpha(\mu)$ clarifies the distinction between attraction and repulsion.
Consider the $C_0$ beta function,
\begin{eqnarray}
\beta (C_0) &=& \mu \frac{d}{d\mu} C_0(\mu) \ =\ \frac{M}{2\pi} C_0(\mu)^2 \ .
\label{eq:rg1}
\end{eqnarray}
Solving this equation gives the RG evolution,
\begin{eqnarray}
  \alpha(\nu) &=& \frac{\alpha(\mu)}{1-\alpha(\mu)\log{\left(\frac{\nu}{\mu}\right)}} \ .
\label{eq:rg1sol}
\end{eqnarray}
In the attractive case, $\alpha(\nu)=-|\alpha(\nu)|$, and the coupling
is asymptotically free, whereas in the repulsive case,
$\alpha(\nu)=|\alpha(\nu)|$, and the coupling increases monotonically
with scale until it hits the Landau pole at $\nu=\mu
\exp\left({1}/{\alpha(\mu)}\right)$ which coincides with the binding
momentum. This is illustrated in Fig.~\ref{fig:alpharun} with the
choice $\mu=1$ and $|\alpha(1)|=1$\footnote{Note that here natural units are chosen such that $\mu=1$ corresponds to a typical infrared physical scale of the system.}. One sees that in the repulsive case, the 
singularity of the $S$-matrix coincides with the position of the
Landau pole, which marks the upper limit of the perturbative description in
terms of $\alpha$, and is therefore unphysical. The physical cutoff of the
EFT is therefore determined by the smaller of this scale and the scale $R^{-1}$
which characterizes the range of the interaction. In the attractive case,
the universal interaction is UV complete, and the EFT is valid below the scale $R^{-1}$.

\subsection{Scattering in the EFT: p-wave}
\label{sec:pwavescatt}
\begin{figure}[!h]
\centering
\includegraphics[width = 0.42\textwidth]{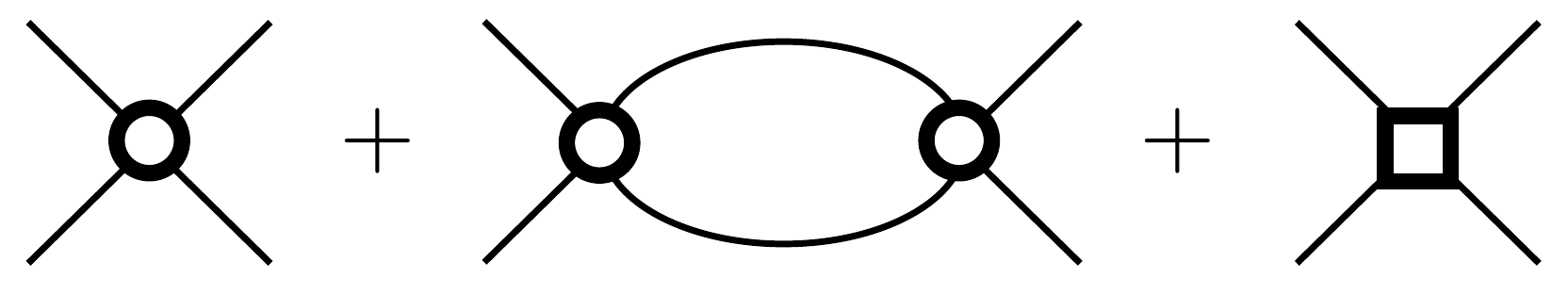}
\caption{Leading p-wave contributions to scattering. The empty circle (square) corresponds to an insertion of the $C'_2$ ($C'_4$) operator.}
    \label{fig:scattp}
\end{figure}

\noindent The contribution to the p-wave scattering amplitude up to next-to-leading order (NLO) is given by the sum of Feynman diagrams shown in Fig.~\ref{fig:scattp} which give
\begin{eqnarray}
  T_1(k,\phi) & = & - C'_2\,{\bf k}\cdot{\bf k}' \ -\ C^{\prime 2}_2 M k_i'k_j \left ( \frac{\mu}{2}\right )^{\epsilon} \int \!{{{\rm d}}^{d-1}{\bf  q}\over (2\pi)^{d-1}}\, 
  \left({{q_i q_j}\over k^2  -{q}^2 + i\delta}\right) \ 
  - \  C'_4\,k^2 {\bf k}\cdot{\bf k}' \nonumber \\
& = &  {\bf k}\cdot{\bf k}'\bigg\lbrack - C'_2 \ -\ C^{\prime 2}_2 {\small \frac{1}{2-\epsilon}} I_1(k) \ -\  C'_4\,k^2 \bigg\rbrack\ .
  \label{eq:T1pwave}
\end{eqnarray}
Defining the renormalized coefficient, $C'_4(\mu)$, with $\overline{MS}$ gives
\begin{eqnarray}
C'_4 & \equiv & C'_4(\mu)\ -\  C^{\prime 2}_2 \frac{M}{8\pi} \bigg\lbrack \ \gamma \ - \log\pi\ -\ \frac{2}{\epsilon} \ - 1 \bigg\rbrack \ ,
  \label{eq:MSbarC4}
\end{eqnarray}
and the renormalized scattering amplitude is then
\begin{eqnarray}
T_1(k,\phi)  & = & {\bf k}\cdot{\bf k}'\bigg\lbrack - C'_2 \ -\ C^{\prime 2}_2 \frac{M k^2}{8\pi} \log{\left(-\frac{k^2}{\mu^2}\right)} \ -\  C'_4(\mu)\,k^2 \bigg\rbrack\, .
  \label{eq:T3Bpwave}
\end{eqnarray}
Now matching to the scattering amplitude of Eq.~(\ref{eq:pwaveSApt}) gives
\begin{eqnarray}
\sigma_p &=& \frac{M C'_2}{8} \ \ \ , \ \ \ \alpha_p(\mu) \ =\  \frac{4\pi C'_4(\mu)}{M C^{\prime 2}_2} \ ,
\label{eq:match1apwave}
\end{eqnarray}
where $a_p = \mu^{-1}\exp{(\alpha_p(\mu))}$.

A noteworthy feature of the p-wave interaction, which is also the case in three dimensions, is that
if the leading operators $C_2'$ and $C_4'$ are treated to all orders, then subleading counterterms are required~\cite{Bertulani:2002sz}.
The highly-singular nature of the p-wave interaction renders the all-orders renormalization subtle, and analogous
to all-orders renormalization of range corrections in the s-wave~\cite{Beane:2021dab}. In this paper, only the leading order
in the perturbative expansion of the p-wave scattering amplitude will be considered. An important distinction from the s-wave is that
the leading p-wave effect, due to $C_2'$, does not run with the RG in $\overline{MS}$ when one-loop effects are included.

\section{Finite-density technology}
\label{sec:FDT}

\subsection{Ideal gas and in-medium modifications}

\noindent In two dimensions, the density $\rho=N/A$, with $N$ the number of particles and $A$ the spatial
area enclosing the particles, of a noninteracting
system with Fermi momentum $\kf$ and degeneracy $g$ is
\beq
\rho\ =\ g \int \! \frac{d^2 {\bf k}}{(2\pi)^2} \theta(\kf-k)
      \ =\  \frac{g\,\kf^2}{4\pi}\ .
\label{fermiden2d}
\eeq
Here, a common Fermi momentum is considered for all spin components, that is, a spin-balanced Fermi gas.
With free single-particle energy $\omega_{\bf k}\equiv {\bf k}^{\,2}/(2M)$, 
the energy density ${\cal E}_0$ of a noninteracting Fermi gas is 
\beq
       {\cal E}_0 \ = \  g \int \! \frac{d^2 {\bf k}}{(2\pi)^2}\,\omega_{\bf k}\,\theta(\kf-k) \ = \ \rho\,\frac{1}{2}\,\frac{\kf^2}{2M}\ .
\label{eden0FG}
       \eeq
The energy per particle is $E/N={\cal E}_0 / \rho$, and can be written as
\beq
E/N \ =\  {\varepsilon}_{FG} \ =\  \frac{\kf^2}{4M} \ =\  \frac{1}{2}\,\varepsilon_F  \ ,
\label{eoNFG}
       \eeq
with $\varepsilon_F \equiv \kf^2/(2M)$.

Feynman diagrams can be used to compute the effect of interactions on
the energy density, and the relevant Feynman rules can be found in
Refs.~\cite{Hammer:2000xg, Furnstahl:2006pa}. In particular, the
interaction vertices can be read off of Eq.~(\ref{lag}) and, for
corrections to the energy at weak coupling, internal lines
are assigned propagators $iG_0 (\kt)_{\alpha\gamma}$, where $\kt
\equiv (k_0,{\bf k})$ is the three-momentum assigned to the line,
$\alpha$ and $\gamma$ are spin indices, and
\begin{align}
    iG_0 (\kt)_{\alpha\gamma}\ =\ iG_0 (\kt)\delta_{\alpha\gamma} &=\ i\delta_{\alpha\gamma}
    \left( \frac{\theta(k-\kf)}{k_0-\omega_{\bf k}+i\delta}
      +\frac{\theta(\kf-k)}{k_0-\omega_{\bf k}-i\delta}\right) \nonumber \\
      &= \ \delta_{\alpha \gamma} \left (\frac{i}{k_0 - \omega_{{\bf k}} +i \delta} - 2 \pi  \delta(k_0 -\omega_{{\bf k}})\theta(\kf - k) \right ) \ .
      \label{freeprop}
\end{align}
The second line breaks the propagator into ``free" and ``in-medium" components.

\subsection{In-medium counting scheme}

In medium, the free-space power counting of Eq.~(\ref{eq:kpowercount}) gets simply modified by setting $E=0$ which gives
\begin{equation}
\chi \ = \   d+1 + 
        \sum_{n=2}^\infty \sum_{i=0}^\infty \left(2i+(d-1)n-d-1\right) V_{2i}^n  \ .
        \label{eq:kfpowercount}
\end{equation}
This contributes to the energy density at order $\kf^\chi R^{\chi-d-1}$ where the powers of $R$ follow from dimensional analysis. 
For universal interactions ($C_0$ only), $\chi=d+1+(d-3)V_0^2$. Therefore, in three dimensions, $\chi=5+V_0^2$ and there is a perturbative expansion with a new power of $\kf R$ accompanying each insertion of $C_0$. By contrast,
in two dimensions, $\chi=4$, and the energy density is given by
\begin{equation}
{\cal E} \ =\ \kf^4\,f(\alpha) \ ,
        \label{eq:edensscaling}
\end{equation}
where $f$ is a function of $\alpha$ such that ${\cal E}_0= \kf^4\,f(0)$. In the Fermi liquid regime
considered here, $f$ admits a power series expansion in $\alpha$, and the goal is to compute
\begin{equation}
{\cal E}_{FL} \ =\ \sum_{n=0}^{n_{max}} {\cal E}_{n}
        \label{eq:flexp}
\end{equation}
up to $n_{max}$. This work will consider $n_{max}=3$ corresponding to three orders in $\alpha$.

\subsection{Thermodynamic potential and superfluid gap}
\label{sec:gap}

\noindent The EFT and power counting outlined above apply to the two-dimensional Fermi gas with weak repulsive coupling $\alpha$.  In the
presence of an arbitrarily weak attractive interaction, the BCS
mechanism causes the Fermi surface to become unstable.  This leads to
pairing superfluidity (superconductivity) for neutral (charged)
fermions which spontaneously breaks the particle-number symmetry
through the formation of a gap, or
condensate~\cite{Polchinski:1992ed,Shankar}.  In making comparisons with
numerical simulations in the attractive regime, it is necessary to
subtract the contribution to the energy density that arises from the
presence of the superfluid gap\footnote{Note that a unified EFT
treatment of the weakly-attractive Fermi liquid has been developed in
Ref.~\cite{Furnstahl:2006pa}.}.

The superfluid gap in
two dimensions was originally computed in
Refs.~\cite{PhysRevLett.62.981,PhysRevB.41.327}.  Here the s-wave gap
in two dimensions is computed in the $\overline{MS}$ scheme in two
ways: by a direct construction and minimization of the renormalized
thermodynamic potential, following Ref.~\cite{LOKTEV20011}, and via a
direct solution of the self-consistent gap
equation~\cite{Papenbrock:1998wb,Marini_1998,Schafer:2006yf,Heiselberg:2000ya}.

As long as $C_0<0$, it is necessary to treat the interactions which
are kinematically enhanced by the BCS mechanism to all orders, in
direct violation of the power-counting rules introduced above. For
consideration of pairing phenomena, it is convenient to view the EFT
somewhat more expansively.  Beginning
with the effective Lagrangian defined in Eq.~(\ref{lag}), with
universal interactions only and $g=2$, one goes to Euclidean space via
$t\rightarrow -i \tau$, and ${\cal L}\rightarrow-{\cal L}_E$ to give
\beqa
  {\cal L}_E  &=&
       \psi^\dagger \bigg\lbrack \partial_\tau - \frac{\nab^{\,2}}{2M}-\mu_F \bigg\rbrack
                 \psi + \frac{C_0}{2}(\psi^\dagger \psi)^2 \ ,
\label{lagE}
\eeqa
and a chemical potential, $\mu_F$, has been introduced for $\psi$ (not to be confused
with the DR scale $\mu$). The partition function is then
\beqa
  {\cal Z}  &=& \int \! {{\cal D} \psi}{{\cal D} \psi^\dagger} \exp\Big\lbrack -\int \! d^3x   {\cal L}_E\Big\rbrack \ .
\label{partf1}
\eeqa
Now if ${\cal Z}$ can be computed at finite temperature $T$, then the thermodynamic potential
is known and given by
\beqa
\beta\,\Omega\left(A,\mu_F,T \right)  &=& -\ln {\cal Z} \left(A,\mu_F,T \right) \, ,
\label{thermopot1}
\eeqa
where $A$ is the area and $\beta\equiv 1/T$ with $k_B=1$. 
The solution can be found by introducing a complex auxiliary field $\Phi = C_0 \psi_{\uparrow} \psi_{\downarrow}$ which decouples the four-Fermi interaction and whose expectation value gives the superfluid gap. 
The Euclidean action is now bilinear in the fermion fields and is formally solved in terms of a fermionic determinant.

Neglecting fluctuations in the fields, that is assuming $\Phi=\text{constant}\neq 0$,
gives the bare thermodynamic potential at zero temperature~\cite{LOKTEV20011}
\beq
\Omega\left(A,\mu_F,\Phi,\Phi^* \right) \ =\  A\,\Big\lbrack  - \frac{1}{C_0}|\Phi|^2 \ -\ \int \! {d^2 {\bf q}\over (2\pi)^2}
\left(  \sqrt{\left(\omega_{\bf q}-\mu_F\right)^2 + |\Phi|^2} - \left(\omega_{\bf q}-\mu_F\right) \right)\Big\rbrack \ .
\label{omegaeff2}
\eeq
These divergent integrals may be evaluated with DR using the formula\footnote{Note that one may take a derivative of the integral with respect to $|\Phi|$, evaluate using DR and then integrate
with respect to $|\Phi|$.}
\begin{eqnarray}
\!\!\!\!\!{\tilde I}(\Phi) &=& \left({\mu\over 2}\right)^{\epsilon} \int \! {d^{d-1} {\bf q}\over   (2\pi)^{d-1}}
{1 \over \sqrt{\left(\omega_{\bf q}-\mu_F \right)^2 + |\Phi|^2}} \nonumber \\[4pt]
&=& {\frac{M}{2 \pi} \left [\frac{2}{\epsilon} + \log \left(\frac{\mu^2 \pi}{M \mu_F}\right ) -\gamma - \log \left (\sqrt{1+|\Phi|^2/\mu^2_F}-1 \right ) \right ]} \ .
\label{eq:Itildegapraw}
\end{eqnarray}
Renormalizing with $\overline{MS}$ using Eq.~(\ref{eq:MSbar}), and exchanging the renormalized
coupling for the two-body binding energy using Eq.~(\ref{eq:bmswave}) then gives the renormalized
thermodynamic potential~\cite{LOKTEV20011}
\beq
\Omega\left(A,\mu_F,\Phi,\Phi^* \right) \ =\  A \frac{M}{4\pi} |\Phi|^2 \Bigg\lbrack
\log \frac{\sqrt{\mu^2_F+|\Phi|^2}-\mu_F}{|\varepsilon_B|} \ -\ \frac{\mu_F}{\sqrt{\mu^2_F+|\Phi|^2}-\mu_F} \ -\ \frac{1}{2}
\Bigg\rbrack \ .
\label{omegaeffinal}
\eeq
This remarkable formula immediately reveals that the superfluid state is energetically favorable and is intrinsically
related to two-body binding~\cite{PhysRevLett.62.981,PhysRevB.41.327}. The minimum of the potential occurs at $\Phi=\Phi^*=\Delta_{LO}$ and defines the leading-order
gap
\begin{eqnarray}
\Delta^2_{{\rm LO}} & = & \varepsilon_B^2\,+\, 2\mu_F|\varepsilon_B| \ .
\label{eq:gaprenormexact}
\end{eqnarray}  
The density is given by
\begin{equation}
\rho  \ = \ -\frac{1}{A}\frac{\partial \Omega(A,\mu_F,\Phi,\Phi^*)}{\partial \mu_F} \ = \ \frac{M}{2\pi}\left ( \mu_F + \sqrt{\Delta_{{\rm LO}}^2 + \mu_F^2} \right )
\label{eq:gapdensity}
\end{equation}
and, after using Eq.~(\ref{fermiden2d}), results in
\begin{eqnarray}
2\, \varepsilon_F & = & \mu_F \ +\ \sqrt{\mu_F^2 \,+\, \Delta^2_{{\rm LO}}} \ .
\label{eq:densrenormexact}
\end{eqnarray}  
Combining Eq.~(\ref{eq:gaprenormexact}) and Eq.~(\ref{eq:densrenormexact}) gives finally, for $\mu_F>0$,
\begin{eqnarray}
\Delta_{{\rm LO}} & = & \sqrt{2\,\varepsilon_F|\varepsilon_B|} \ \ \ \ , \ \ \ \  \mu_F \ = \ \varepsilon_F\ -\ \frac{1}{2}|\varepsilon_B| \ ,
\label{eq:gapfinal}
\end{eqnarray}  
in agreement with Ref.~\cite{PhysRevLett.62.981,PhysRevB.41.327}.

\begin{figure}[!ht]
\centering
\includegraphics[width = 0.6\textwidth]{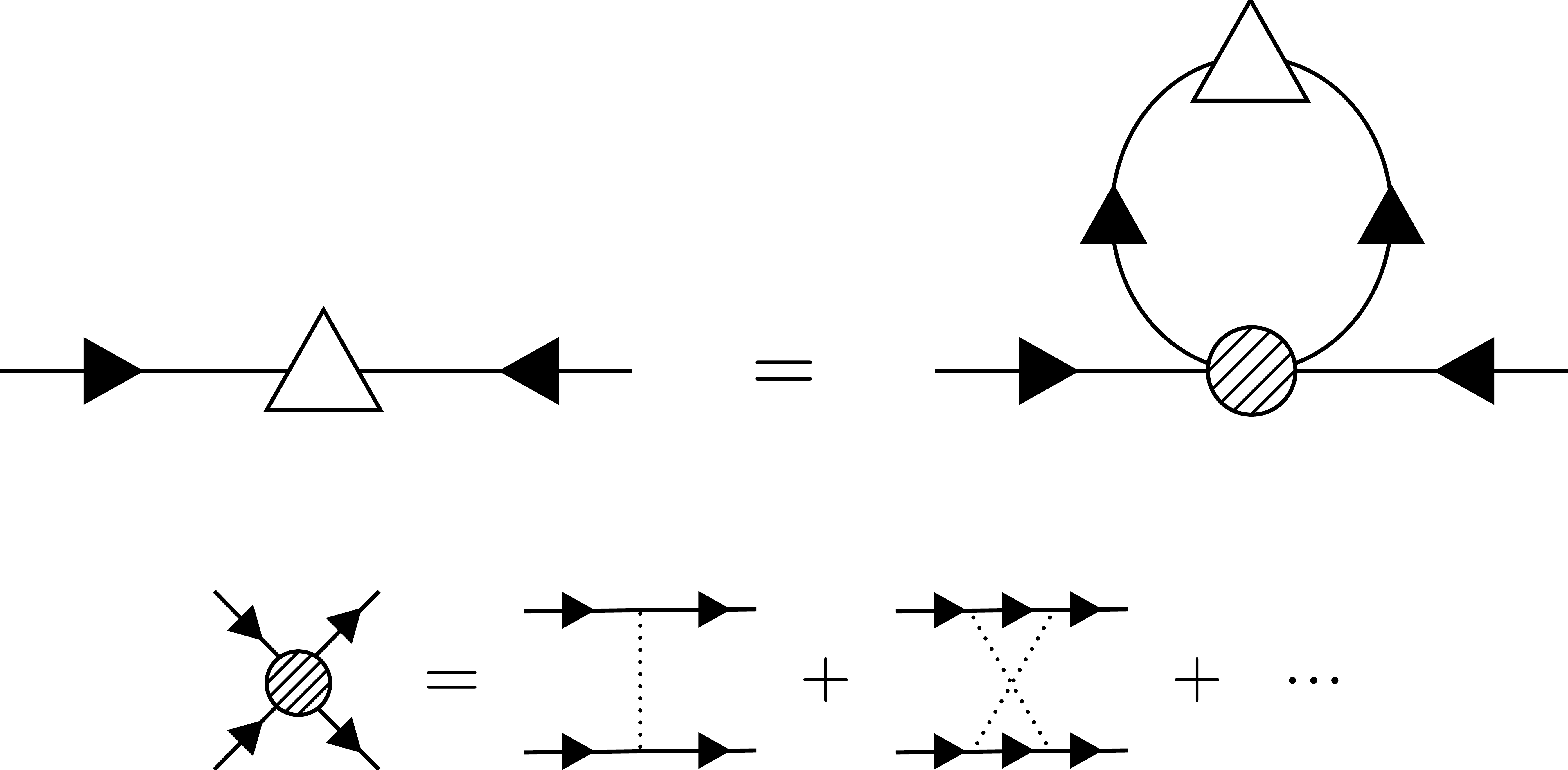}
\caption{The gap equation diagrammatically (top), with the empty triangle denoting an insertion of the gap. The shaded circle represents an insertion of the
two-particle irreducible potential $-i V_{pp}$ (bottom), with the dotted lines denoting a $C_0$ interaction. Crossed diagrams and diagrams which cancel are not shown.}
    \label{fig:GAP}
\end{figure}
It is instructive to obtain this result using Feynman diagrams as they
render in-medium corrections more transparent. The gap equation is shown
diagrammatically in Fig.~\ref{fig:GAP} where the $C_0$ vertex is represented by a dotted line to distinguish which fermion lines are being contracted~\cite{alma991009084709703276}. This is evaluated to LO by
taking the two-particle irreducible potential equal to the tree-level
contact vertex, $V_{pp}=C_0$, and using a propagator which accounts
for the gap~\cite{Schafer:2006yf,LOKTEV20011}. The result is
\begin{eqnarray}
\Delta_{{\rm LO}}  & = & -C_0\int \! {d^3 {q}\over   (2\pi)^3}
{\Delta_{{\rm LO}} \over q_3^2+\left(\omega_{\bf q}-\mu_F \right)^2 + \lvert \Delta_{{\rm LO}}\rvert ^2}
\ ,
\label{eq:gap}
\end{eqnarray}
where $q_3$ is the third component of Euclidean momentum.
The gap is the self-consistent solution to this equation, which treats
$C_0$ to all orders even if it is arbitrarily weak because of the
kinematical enhancement (BCS instability) of the loop function for
$|{\bf q}|\sim \kf$.  Performing the $q_3$ integration leaves
\begin{eqnarray}
-\frac{1}{C_0} & = & \frac{1}{2} {\tilde I}\left( \Delta_{{\rm LO}} \right)
\ .
\label{eq:gapraw}
\end{eqnarray}
Using Eq.~(\ref{eq:Itildegapraw}) and renormalizing with $\overline{MS}$ using Eq.~(\ref{eq:MSbar}) then gives
\begin{equation}
\frac{1}{C_0(\sqrt{M \mu_F})}= \frac{M}{4\pi}\log\left (\sqrt{1+\frac{\Delta_{LO}^2}{\mu_F^2}} -1 \right )
\label{eq:gaprenorm}
\end{equation}  
which immediately recovers Eq.~(\ref{eq:gaprenormexact}).

The energy density of the paired state is given by
\begin{equation}
{\cal E}   =  \ A^{-1} \Omega \ + \ \mu_F \rho \, = \ {\cal E}_0 \ -\ \frac{M}{4\pi}\Delta_{LO}^2 \ .
\label{eq:gapenergydensity2}
\end{equation}
In the literature this approximation of the energy is denoted mean-field BCS theory.
The contribution of the gap to the energy-per-particle is thus given by~\cite{Gorkov_ContributionTheorySuperfluidity_1961,Heiselberg:2000ya,PhysRevA.67.031601,Resende:2012zs,Schafer:2006yf}
\beq
(E/N)_\Delta \ =\   \frac{1}{2}{\varepsilon}_B \ .
\label{eppgc}
\eeq
Therefore, the existence of a bound state is a necessary and
sufficient condition for the existence of s-wave pairing, in contrast
to the three dimensional
case~\cite{PhysRevLett.62.981,PhysRevB.41.327}. At weak, attractive
coupling, the contribution of the gap to the energy is exponentially
suppressed, which, as will be seen, allows a meaningful perturbative
expansion in $\alpha$.

The NLO contribution to the gap takes into account the particle-hole
(``ring") correction to the two-particle irreducible potential
$V_{pp}$~\cite{Gorkov_ContributionTheorySuperfluidity_1961,Heiselberg:2000ya,PhysRevA.67.031601,Resende:2012zs,Schafer:2006yf}
as shown in the bottom of Fig.~\ref{fig:GAP}.  This accounts for the
polarizibility of the finite-density medium which effectively screens
the contact interaction. For the kinematics which lead to the BCS
instability, ${\bf k}_1 = -{\bf k}_2 \equiv {\bf k}$, ${\bf k}_1' =
-{\bf k}_2' \equiv {\bf k}'$ and $k = k' = \kf$, the potential may be computed
in the gapless EFT with $\kf R \ll 1$ to give
\begin{equation}
    V_{pp} = C_0(\delta_{\alpha \gamma}\delta_{\beta\delta}-\delta_{\alpha \delta}\delta_{\beta\gamma}) \ + \  i C_0^2 \int \! \frac{d^3 q}{(2\pi)^3} \left [ G_0(\tilde{q}) \, G_0(\tilde{q} + \tilde{P}_+)\delta_{\alpha \gamma}\delta_{\beta \delta} - G_0(\tilde{q}) \, G_0(\tilde{q} + \tilde{P}_-) \delta_{\alpha \delta}\delta_{\beta \gamma}\right ]
\end{equation}
where $\tilde{P}_{\pm} = (0, {\bf k}\pm {\bf k}')$. 
Spin indices have temporarily been restored and $\alpha, \beta$ ($\gamma, \delta$) are the spin indices of the incoming (outgoing) particles.
To project the potential onto the s-wave one must integrate over all $\cos\theta = \hat{\bf k}\cdot \hat{\bf k}'$. 
However, it is straightforward to show that $V_{pp}$ computed to one loop contains no partial waves higher than $\ell = 0$~\cite{PhysRevB.48.1097}.
Therefore the two terms with $\tilde{P}_{\pm}$ contribute equally to $V_{pp}$ which becomes
\begin{equation}
V_{pp} \ = \ C_0 \ + \ 2 M C_0^2\int\displaylimits_{q<\kf}\! \frac{d^2 {\bf q}}{(2\pi)^2}\bigg [\frac{1}{P_-^2/2+ {\bf q}\cdot {\bf P}_- - i \epsilon} + i \pi \delta(P_-^2/2 + {\bf q}\cdot {\bf P}_-) \theta(\kf - \lvert {\bf q} + {\bf P}_- \rvert) \bigg ] \ ,
\end{equation}
where spin indices have again been suppressed.
Evaluating the integral\footnote{This integral appears in the NNLO correction to the energy density and is evaluated below, see Eq.~(\ref{eq:IRing}).} results in
\begin{equation}
    V_{pp} \ = \ C_0 + \frac{M}{2\pi}C_0^2 \ .
\end{equation}
The gap equation then becomes
\begin{eqnarray}
-\frac{1}{C_0}\left( 1+\frac{M}{2\pi}C_0\right)^{-1} & = & 
-\frac{1}{C_0} \ +\ \frac{M}{2\pi} \ +\ {\it O}(C_0)\ =\
\frac{1}{2} {\tilde I}\left( \Delta_{{\rm NLO}} \right)
\ .
\label{eq:gaprawNLO}
\end{eqnarray}
Neglecting the ${\it O}(C_0)$ corrections on the left hand side
then leads to the gap energy up to NLO
\begin{equation}
   \Delta_{{\rm NLO}} = \frac{1}{e}\Delta_{{\rm LO}} \ ,
\label{eq:dnlo}
\end{equation}
in agreement with Ref.~\cite{PhysRevA.67.031601}.  

As the omitted corrections to $V_{pp}$ are ${\it O}( C_0^3 )$, one expects that
Eq.~(\ref{eq:dnlo}) is valid up to corrections of ${\it O}(
\alpha \Delta_{{\rm NLO}})$.  It is important to stress that while the
computation of the gap is valid for all interparticle separations
$\kf^{-1}$, the EFT giving rise to this screening correction is
strictly valid at large interparticle separations. Indeed, at strong
coupling, the paired fermions are expected to become tightly bound,
leading to the BCS-BEC crossover to a gas of repulsive
bosons. Clearly, in this limit, screening effects will become
negligible as the diameter of the pair will be much smaller than
$\kf^{-1}$, and therefore it is expected that the LO gap contribution
to the energy per particle, ${\varepsilon}_B/2$, will be
exact. Notice, however, that mean-field BCS theory in the strong
coupling molecular limit misses the correct interaction energy between
composite bosons~\cite{PhysRevA.67.031601,BertainaS}.  In particular,
the first term in Eq.~(\ref{eq:gapenergydensity2}) is not correct in
the BEC limit because it should include the interaction energy of the
composite bosons.

\section{Weakly-coupled Fermi gas: universal corrections}
\label{sec:DFGu}

\subsection{Fermi liquid regimes}

\noindent The goal in what follows is to compute the energy density
in the weak coupling, Fermi liquid regime, $|\alpha|\ll 1$, which
has been shown to divide into a repulsive and an attractive branch as
\begin{eqnarray} 
{\cal E} \ =\ \,\begin{cases}
{\cal E}_{FL}            \ , \qquad\qquad\qquad\quad\,\,\; \alpha > 0\\
{\cal E}_{FL} \ -\ \frac{M}{4\pi}\Delta^2 \ , \ \qquad\quad \alpha < 0\ ,
\end{cases}
\label{Smatmobius1a}
  \end{eqnarray}
where $\Delta=\Delta_{{\rm NLO}}(1+ {\it O}(\alpha))$. 
Note that although the energy due to pairing is exponentially small in $\lvert \alpha \rvert$, it is included in order to be able to consistently compare with the Monte Carlo simulations. 
In what follows the perturbative calculation of ${\cal E}_{FL}$ in powers of $\alpha$
is described order-by-order.

\subsection{Leading order (LO)}

\begin{figure}[!h]
\centering
\includegraphics[width = 0.24\textwidth]{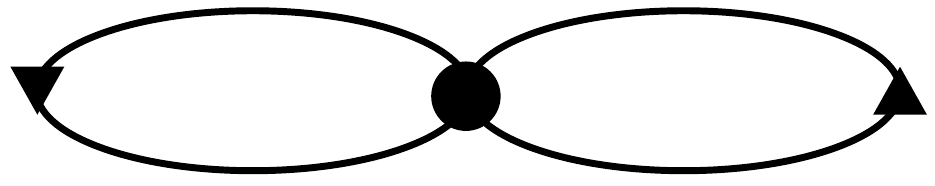}
\caption{Leading order diagram (the bow tie) contributing to the energy density. The black circle corresponds to an insertion of the
  $C_0$ operator.}
    \label{fig:EMF}
\end{figure}

\noindent The LO diagram which contributes to the energy density, ${\cal E}_{FL}$, is shown in Fig.~\ref{fig:EMF} and yields
\beq
 {\cal E}_1 = \frac{1}{2}\,C_0\,g(g-1)\left(
    \lim_{\eta\to 0^+} \int \! \frac{d^3 k}{(2\pi)^3}\,e^{ik_0 \eta}\,iG_0(\kt)
    \right)^2 \ .
\eeq
The $dk_0$ integration is performed using contour integration, which
picks up the $\theta(\kf-k)$ pole, and the remaining $d^2 {\bf k}$ integral
up to $\kf$ is trivial.  The result is
\beq
  {\cal E}_1 =\rho\,(g-1)\,\frac{\kf^2}{8\pi}\,C_0 \ ,
\eeq
and is sometimes referred to as the mean-field
contribution. 
Using Eq.~(\ref{eq:MSbarpert}) to replace the bare coupling with the renormalized coupling, and using Eq.~(\ref{eq:9b}) to express the final result in terms of $\alpha$, it is found that
\beqa
    {\cal E}_1  &=& \rho (g-1) \frac{\kf^2}{4M} \bigg\lbrack  \alpha(\mu) \ +\ {\it O}(\alpha(\mu)^2 )  \bigg\rbrack \ .
   \label{energyCmf}
   \eeqa
At this order, the RG scale is arbitrary and will be set once the NLO contributions are taken into account.
The factor of $g-1$ will be common to all universal corrections and reflects that this interaction must vanish for single-component fermions due to Pauli statistics.

\subsection{Next-to-leading order (NLO)}
\begin{figure}[!h]
\centering
\includegraphics[width = 0.46\textwidth]{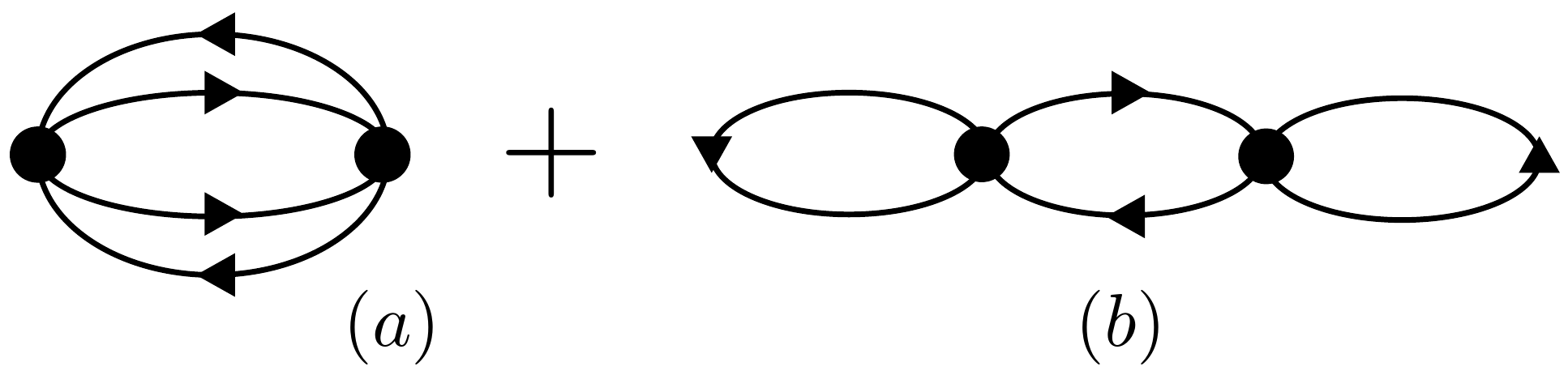}
\caption{Next-to-leading order diagrams contributing to the energy density. Only diagram (a), the beach ball diagram, is non-vanishing.}
    \label{fig:ENLO}
\end{figure}
\noindent The NLO calculations in this section and the NNLO calculations in the following section reproduce the results first found in Refs.~\cite{Kaiser:2013bua,Kaiser:2014laa}.
The nominal NLO corrections come from the diagrams in Fig.~\ref{fig:ENLO}. Diagram $(b)$ is ``anomalous" and is identically zero. 
 The contribution from diagram $(a)$, the beach ball diagram, is
\beq
  \label{eq:energy2a}
   {\cal E}_2 = -i\,\frac{C_0^2}{4}\,g(g-1)
    \int \! \frac{d^3 p_1}{(2\pi)^3}\int \! \frac{d^3 p_2}{(2\pi)^3}
    \int \! \frac{d^3 k}{(2\pi)^3}\,
	      G_0(\tilde{p}_1)\,G_0(\tilde{p}_2)\,G_0(\tilde{P}+\tilde{k})\,G_0(\tilde{P}-\tilde{k}) \ ,
\eeq
where $\tilde{P} = (\tilde{p}_1 + \tilde{p}_2)/2$ and it is convenient to define $\tilde{q} = (\tilde{p}_1 - \tilde{p}_2)/2$. 
Using the in-medium form of the propagator, it is found that only terms with two in-medium insertions on either side of a loop survive the contour integration.
Without loss of generality, the two in-medium insertions may be placed on the $\tilde{p}_{1,2}$ loop.
This puts $\tilde{p}_{1,2}$ on shell and restricts their momenta to be below the Fermi surface.
After performing the contour integrals, the energy becomes
\begin{align}
  \label{eq:energy2acont}
   {\cal E}_2 = \frac{MC_0^2}{4}\,g(g-1)
    \int\displaylimits_{p_{1,2} < \kf}\!\frac{d^2 {\bf p}_1 d^2 {\bf p}_2}{(2\pi)^4}\,
	\bigg [2{\cal I}_0 + 2{\cal I}_1 + {\cal I}_2 \bigg ] \ ,
\end{align}
where
\begin{align}
    \label{eq:intDef}
    {\cal I}_0 &= \ \left (\frac{\mu}{2}\right )^{\epsilon}\int\! \frac{d^{d-1} {\bf k}}{(2\pi)^{d-1}} \frac{1}{q^2-k^2 +i \delta} \ , \\[4pt]
    {\cal I}_1 &= \ -\int\! \frac{d^2 {\bf k}}{(2\pi)^{2}} \frac{\theta(\kf-|{\bf P}+{\bf k}|)+\theta(\kf-|{\bf P}-{\bf k}|)}{q^2-k^2+i\delta} \ , \\[4pt]
    {\cal I}_2 &= \ -2 i \pi \int\! \frac{d^2 {\bf k}}{(2\pi)^{2}}\delta(k^2 - q^2) \theta(\kf-\lvert {\bf P} - {\bf k} \rvert) \theta(\kf-\lvert {\bf P} + {\bf k} \rvert) \ ,
\end{align}
and $\overline{MS}$ is used to define ${\cal I}_0$ ($=I_0(q)/M$). The energy is manifestly real and, for $p_{1,2}<\kf$,
\begin{equation}
{\rm Im} \, \left (2{\cal I}_0 + 2{\cal I}_1 + {\cal I}_2 \right ) = 0 \ .
\end{equation}
After changing to dimensionless variables, $s=P/\kf$ and $t = q/\kf$, the real parts are found to be
\begin{align}
    \label{eq:intRe}
    {\rm Re} \, {\cal I}_0 &= \frac{1}{4\pi} \Bigg\lbrack \log{\left(\frac{t^2 \kf^2}{\mu^2}\right)}\ +\ \gamma \ - \log\pi\ -\ \frac{2}{\epsilon} \Bigg\rbrack \nonumber \ , \\
    {\rm Re} \, {\cal I}_1 &= \ -\frac{1}{2\pi} \Bigg\lbrack \log{t}\ -\ H(s,t) \Bigg\rbrack \ ,
\end{align}
where~\cite{Kaiser:2013bua}
\begin{align}
    H(s,t) &=   2\,\theta(1-s-t)\,\log\frac{\sqrt{1-(s+t)^2}+\sqrt{1-(s-t)^2}}{2\sqrt{t}} \  +\ \theta(s+t-1)\,\log{s} \ .
   \label{eq:hdef}
\end{align}
After integrating by parts, the following identity is obtained
\begin{align}
      & \int\displaylimits_{p_{1,2} < \kf}\!\frac{d^2 {\bf p}_1 d^2 {\bf p}_2}{(2\pi)^4} f(s,t) \ = \ \frac{2\kf^4}{\pi^3} \int\displaylimits_{0}^1 \! ds \, s \int\displaylimits_{0}^{\sqrt{1-s^2}} \! dt \, t \, J(s,t) f(s,t)
   \label{eq:Jdef}
\end{align}
where $f(s,t)$ is an arbitrary function and
\begin{equation}
    J(s,t) =   \frac{\pi}{2} \theta(1-s-t)\ +\ \theta(s+t-1)\,\arcsin \frac{1-s^2-t^2}{2 s t} \ .
   \label{Jdef}
\end{equation}
Applying this to Eq.~\ref{eq:energy2acont}, one finds
\begin{align}
   {\cal E}_2 = M C_0^2\,g(g-1) \frac{\kf^4}{\pi^3}
    \int\displaylimits_{0}^1 \! ds \, s \int\displaylimits_{0}^{\sqrt{1-s^2}} \! dt \, t \, J(s,t) \left [{\rm Re}\, {\cal I}_0 + {\rm Re}\, {\cal I}_1  \right ] \ .
  \label{energy2aSimp}
\end{align}
Notice that the $\log (t)$ term cancels in the integrand and the remaining integration over ${\rm Re}\, {\cal I}_0$ in Eq.~(\ref{energy2aSimp}) gives
\beqa
 \delta{\cal E}_2 &=&  \rho (g-1) \frac{\kf^2}{8\pi}  \frac{C_0^2 M}{4\pi} \Bigg\lbrack \log{\left(\frac{\kf^2}{\mu^2}\right)} \ +\ \gamma \ - \log\pi\ -\ \frac{2}{\epsilon} \Bigg\rbrack
     \ .
   \label{energy2const}
\eeqa
Adding this contribution to the LO energy density, choosing $\mu = \kf$, and replacing the bare parameters with renormalized parameters
using Eq.~(\ref{eq:MSbarpert}), one finds
\beq
  {\cal E}_1\ +\ \delta{\cal E}_2 =\rho\,(g-1)\,\frac{\kf^2}{8\pi}\,C_0(\kf) \ +\ {\it O}( C_0(\kf)^3 )
         \ .
\eeq
The integration over ${\cal I}_1$ in Eq.~(\ref{energy2aSimp}) gives
\beqa
 \delta{\cal E}'_2 \ =\ \rho (g-1) \frac{\kf^2}{16\pi^2}\, MC^2_0\left(\frac{3}{4}-\log 2\right)\, .
   \label{energyslr}
\eeqa
Again, using Eq.~(\ref{eq:MSbarpert}) to renormalize this contribution, and with ${\cal E}_2=\delta{\cal E}_2+\delta{\cal E}'_2$, one finds to NLO
\beqa
    {\cal E}_1 \ +\ {\cal E}_2 &=& \rho (g-1) \frac{\kf^2}{4M}    \Bigg\lbrack  \alpha(\kf) \ +\ \alpha(\kf)^2\left(\frac{3}{4}-\log 2\right)  \ +\ {\it O}(\alpha(\kf)^3 )  \Bigg\rbrack \ ,
   \label{energyCform}
\eeqa
where Eq.~(\ref{eq:9b}) has been used to express the final result in terms of $\alpha$. This recovers the result of Refs.~\cite{PhysRevB.45.10135,PhysRevB.45.12419}.
Note that ${3}/{4}-\log 2=0.05685$ is small as compared to a number of order one. The small size of this correction has been observed in comparison with MC simulations~\cite{PhysRevLett.106.110403}, which
suggest a stronger deviation from the mean-field result, and motivates the study of higher-order effects.

\subsection{Next-to-next-to-leading order (NNLO)}

\begin{figure}[!h]
\centering
\includegraphics[width = 0.85\textwidth]{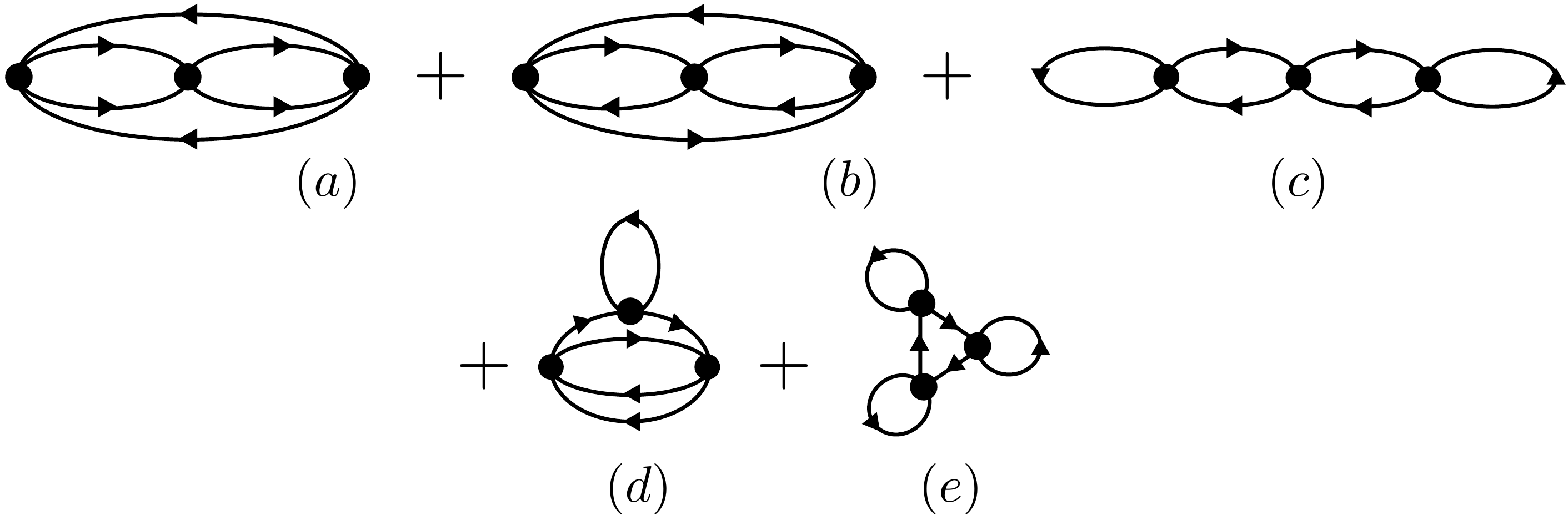}
\caption{Next-to-next-to-leading order diagrams contributing to the energy density. Only diagrams (a) and (b) are non-vanishing.}
    \label{fig:ENNLO}
\end{figure}
\noindent The NNLO corrections come from the diagrams in Fig.~\ref{fig:ENNLO}. 
Diagrams $(c)$, $(d)$ and $(e)$ are all anomalous and evaluate to zero.
The ladder diagram, Fig.~\ref{fig:ENNLO}$(a)$, gives a logarithmically-divergent contribution to the energy
\begin{equation}
     {\cal E}_3^L = g(g-1) \frac{C_0^3}{6} \int \! \frac{d^3 p_1}{(2 \pi)^3} \int \!\frac{d^3 p_2}{(2 \pi)^3} G_0(\tilde{p}_1) \, G_0(\tilde{p}_2)\left [\int \! \frac{d^3 k}{(2 \pi)^3}  G_0(\tilde{P}+\tilde{k}) \, G_0(\tilde{P}-\tilde{k})\right ]^2\ .
\end{equation}
As with the beach ball diagram, only terms with two in-medium
insertions on either side of a loop are non-zero and these will again
be placed on the $\tilde{p}_{1,2}$ loop. Many of the non-vanishing terms
are identical due to the cyclic symmetry of the diagram and the
integrals involved are the same as in the evaluation of the beach
ball. Here, the imaginary parts will need to be kept and the following
relation is particularly useful
\begin{equation}
    {\rm Im}({\cal I}_0 + {\cal I}_1 + {\cal I}_2) = \frac{{\cal I}_2}{2 i} = -\frac{J(s,t)}{2 \pi} \ ,
    \label{eq:imInt}
\end{equation}
which holds for $s^2 + t^2 < 1$.
After performing the contour integration, the energy reads
\begin{align}
  {\cal E}_3^L & =\  g(g-1) \frac{C_0^3}{6} M^2\int\displaylimits_{p_{1,2} < \kf}\!\frac{d^2 {\bf p}_1 d^2 {\bf p}_2}{(2\pi)^4}
  \bigg [3 \left ({\cal I}_0 + {\cal I}_1 +{\cal I}_2\right )^2 - \: {\cal I}_2 \left (3 {\cal I}_0 + 3 {\cal I}_1 +2{\cal I}_2\right )\bigg ] \nonumber \\[5pt]
     & = \frac{\rho (g-1) C_0^3 M^2 \kf^2}{3 \pi^4}\int\displaylimits_{0}^1 \! ds \, s \int\displaylimits_{0}^{\sqrt{1-s^2}}\! dt \, t \, J(s,t)\bigg \{-J(s,t)^2 + 3 H(s,t)^2 \nonumber \\
     &+ \: 3 H(s,t) \left [ \gamma  - \log\pi - \frac{2}{\epsilon} +\log(\kf^2/\mu^2)\right ] + \frac{3}{4} \left [ \gamma  - \log\pi - \frac{2}{\epsilon} +\log(\kf^2/\mu^2)\right ]^2 
      \bigg \} \ ,
 \end{align}
where in the second line, Eqs.~(\ref{eq:intRe}), (\ref{eq:Jdef}) and (\ref{eq:imInt}) have been used.
Replacing the bare coupling with the renormalized coupling in ${\cal E}_1 + {\cal E}_2 + {\cal E}_3^L$, and setting
$\mu = \kf$ to remove the RG scale dependence at this order, results in
\begin{align}
  {\cal E}_3^L & = \ \rho (g-1) \frac{\kf^2}{4M} \bigg [ \alpha(\kf)^3\left(0.16079\right)\ +\ {\it O}(\alpha(\kf)^4) \bigg ] \ ,
\end{align}
where the integrals over $J^3$ and $J H^2$ have been performed numerically.
This is in agreement with the result of Ref.~\cite{Kaiser:2013bua} (Appendix A).

The ring diagram, Fig.~\ref{fig:ENNLO}$(b)$, gives the finite result
\begin{align}
    {\cal E}_3^R= -\frac{i}{6}g(g-1)(g-3) C_0^3 \int \! \frac{d^3 P}{(2\pi)^3} \left [i \int \! \frac{d^3 k}{(2\pi)^3}G_0(\tilde{k})\,G_0(\tilde{k}+\tilde{P}) \right ]^3 \ .
\end{align}
After performing the contour integration, the term in brackets becomes
\begin{align}
    {\cal I}_R = M \int \! \frac{d^2 {\bf k}}{(2\pi)^2}&\bigg [\left (\frac{\theta(\kf-k)}{P^2/2+ {\bf k}\cdot {\bf P} - M P_0 - i \delta} \: + \: P_0 \to -P_0 \right ) \nonumber \\
    &- \: 2 i \pi \delta(MP_0 -P^2/2 - {\bf k}\cdot {\bf P}) \theta(\kf - k) \theta(\kf - \lvert {\bf k} + {\bf P} \rvert) \bigg ] \ .
    \label{eq:IRing}
\end{align}
It is convenient to calculate the real and imaginary parts separately. 
The imaginary part has three terms which, after the change of variables ${\bf k} \to {\bf k} - {\bf P}/2$, give
\begin{eqnarray}
    {\rm Im}\,{\cal I}_R &=& M \pi \int \! \frac{d^2 {\bf k}}{(2\pi)^2} \delta(M P_0 - {\bf k}\cdot {\bf P})\big [\theta(\kf - \lvert {\bf k} - {\bf P}/2 \rvert) + \theta(\kf - \lvert {\bf k} + {\bf P}/2 \rvert) \nonumber \\
    && \quad \quad \quad \quad \quad \ \  - \: 2\theta(\kf - \lvert {\bf k} - {\bf P}/2 \rvert) \theta(\kf - \lvert {\bf k} + {\bf P}/2 \rvert)\big ] \nonumber  \nonumber \\
    &=& \frac{M}{4\pi}I(\bar{\nu},x)
\end{eqnarray}
where~\cite{Kaiser:2014laa}
\begin{align}
&I(\bar{\nu},x) = {1\over x} \sqrt{1-(x-\bar{\nu})^2}\,, \quad  {\rm for} \quad |x-1|<\bar{\nu} <x+1\ , \nonumber \\
&I(\bar{\nu},x) = {1\over x}\bigg [ \sqrt{1-(x-\bar{\nu})^2}-\sqrt{1-(x+\bar{\nu})^2}\bigg ] \,, \quad  {\rm for} \quad 0<\bar{\nu} <1-x\ ,
\end{align}
and dimensionless variables, $x = P/(2 \kf)$ and $2 x \bar{\nu} = M P_0/ \kf^2$, have been defined.
Before calculating the real part, notice that ${\cal E}_3^R$ must be proportional to ${\rm Im}\,({\cal I}_R)^3$ to be real, and therefore will always include at least one factor of $I(\bar{\nu},x)$.
Therefore, ${\rm Re}\,{\cal I}_R$  need only be defined in the semi-infinite strip where $I(\bar{\nu},x)$ has support (Figure 2 in Ref.~\cite{Kaiser:2014laa}). 
In this domain
\begin{equation}
    {\rm Re}\,{\cal I}_R = \frac{M}{4\pi}R(\bar{\nu}, x)
\end{equation}
where
\begin{align}
    & R(\bar{\nu},x) = 2-{1\over x} \sqrt{(x+\bar{\nu})^2-1}\,, \quad  {\rm for} \quad |x-1|<\bar{\nu} <x+1\,, \nonumber \\
    & R(\bar{\nu},x) = 2 \,, \quad  {\rm for} \quad 0<\bar{\nu} <1-x\,,\quad 0<x<1\ .
\end{align}
The energy density is then
\begin{eqnarray}
{\cal E}_3^R &=&-\frac{i}{6} g(g-1)(g-3)C_0^3 \int \! \frac{d^3 P}{(2 \pi)^3}\, ({\cal I}_R)^3 \nonumber \\
&=&\rho (g-1)(g-3) \kf^2 \, \frac{C_0^3 M^2}{24 \pi^4}\int\displaylimits_{0}^{\infty}\! dx \, x^2 \int\displaylimits_{\bar{\nu}_{{\rm min}}}^{x+1} \! d\bar{\nu}
\left [3 R(\bar{\nu},x)^2 I(\bar{\nu},x) -  I(\bar{\nu},x) ^3\right ] \ ,
\end{eqnarray}
where $\bar{\nu}_{{\rm min}} = {\rm max}(0,x-1)$. Evaluating the integral\footnote{The final integration is
  simpler if one uses rotated coordinates, $x = (\xi + \eta)/2 \ ,
  \ \bar{\nu} = (\xi - \eta)/2$ where the integration region is $-\xi < \eta
  < \xi$, $0<\xi<1$ and $-1<\eta<1$, $\xi>1$.}, and setting the RG scale to $\kf$ then gives
\begin{align}
{\cal E}_3^R &=\ \rho (g-1)(g-3)\frac{\kf^2}{4M}  \bigg [\alpha(\kf)^3\left(2 \log{2} - 1\right) \ +\ {\it O}(\alpha(\kf)^4) \bigg ] \ .
\end{align}

Finally, the complete NNLO expression is given by
\begin{align}
  {\cal E}_1 \; +\; {\cal E}_2 \; +\; {\cal E}_3^L \; +\; {\cal E}_3^R & = \ \rho (g-1) \frac{\kf^2}{4M} \bigg [ \alpha(\kf) \; +\; \alpha(\kf)^2\left(\frac{3}{4}-\log 2\right)\nonumber \\
  & +\: \alpha(\kf)^3  \big\lbrack  0.16079 \ + \ (g-3)\left(2 \log 2 -1 \right)\big\rbrack \ +\ {\it O}(\alpha(\kf)^4) \bigg ] \ .
\end{align}
Note that with $g=2$, $0.16079-\left(2 \log 2 -1 \right)=-0.22550$, which is a factor of four larger in magnitude than the $\alpha(\kf)^2$ coefficient.

\section{Weakly-coupled Fermi gas: nonuniversal corrections}
\label{sec:DFGnu}

\subsection{Range corrections}
\begin{figure}[!h]
\centering
\includegraphics[width = 0.25\textwidth]{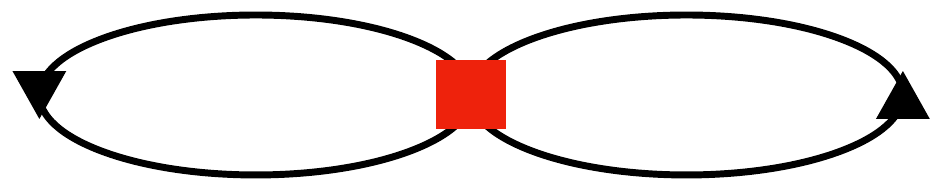}
\caption{Effective range contribution to the energy density. The red square corresponds to an insertion of the $C_2$ operator.}
    \label{fig:ENrangecorr}
\end{figure}
\noindent According to the power-counting formula,
Eq.~(\ref{eq:kfpowercount}), an insertion of the $C_2$ operator gives an
${\it O}(\kf^6)$ contribution to the energy density.  However,
effective range corrections, and indeed corrections from all orders in the
effective range expansion, are also driven by the $C_0$ operator and
therefore will be doubly suppressed in the dilute and weak coupling
limits. From the diagram in Fig.~\ref{fig:ENrangecorr}:
\beqa
    {\cal E}^{\sigma_2}_2 &=& \rho (g-1) \frac{\kf^4}{32\pi} C_2 \ ,
   \label{rangecorr1}
\eeqa
and finally, in terms of renormalized parameters and the two-dimensional effective range defined in Eq.~(\ref{eq:9b}),
\beqa
    {\cal E}^{\sigma_2}_2 &=& \rho (g-1) \frac{\kf^2}{4M}  \alpha(\kf)^2 \frac{\pi}{8} \left(\sigma_2 \kf^2\right) \ .
   \label{rangecorr2}
\eeqa

\subsection{Three-body effects}

\begin{figure}[!h]
\centering
\includegraphics[width = 0.15\textwidth]{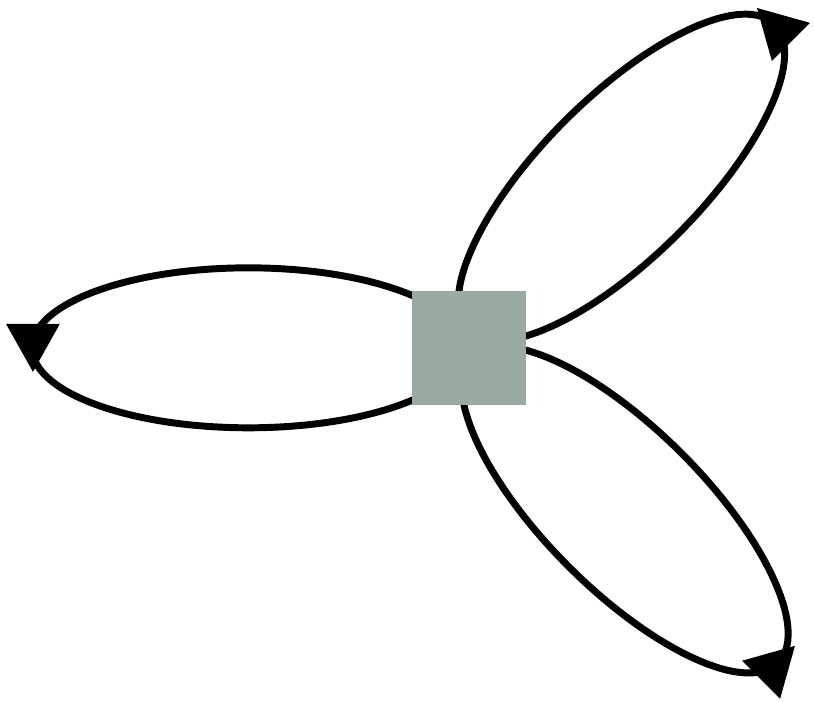}
\caption{Leading three-body contribution to the energy density. The grey square corresponds to an insertion of the $D_0$ operator.}
    \label{fig:EN3bod}
\end{figure}
\noindent From the diagram in Fig.~\ref{fig:EN3bod}:
\beqa
    {\cal E}^{D_0}_0 &=& \rho (g-2)(g-1) \frac{\kf^4}{96\pi^2} D_0 \ ,
   \label{threebod}
\eeqa
and finally
\beqa
    {\cal E}^{D_0}_0 &=& \rho (g-2)(g-1) \frac{\kf^2}{4M} \frac{1}{24\pi^2} \left(M D_0 \kf^2\right) \ .
   \label{threebod2}
\eeqa
This scales with the Fermi momentum like a range correction, but with no additional suppression in $\alpha$.
As a local three-body interaction, it vanishes for $g<3$ due to Pauli statistics.

\subsection{P-wave corrections}
\begin{figure}[!h]
\centering
\includegraphics[width = 0.25\textwidth]{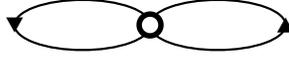}
\caption{P-wave contribution to the energy density. The empty circle corresponds to an insertion of the $C'_2$ operator.}
    \label{fig:ENpwavecorr}
\end{figure}
\noindent From the diagram in Fig.~\ref{fig:ENpwavecorr}:
\beqa
    {\cal E}^{\sigma_p}_0 &=& \rho (g+1) \frac{\kf^4}{32\pi} C'_2 \ ,
   \label{pwavecorr1}
\eeqa
and finally, using Eq.~(\ref{eq:match1apwave}),
\beqa
    {\cal E}^{\sigma_p}_0 &=& \rho (g+1) \frac{\kf^2}{4M}   \frac{1}{\pi} \left(\sigma_p \kf^2\right) \ .
   \label{pwavecorr2}
\eeqa
This again scales with the Fermi momentum like a range correction, but with no additional suppression in $\alpha$, and no longer vanishes for $g=1$ due to the p-wave wavefunction being antisymmetric.

\section{General scale-dependent form and the contact}
\label{sec:summandgen}

\noindent The energy-per-particle of the weakly-coupled Fermi gas in two dimensions including contributions of ${\it O}(\alpha^3)$, ${\it O}(\kf^2)$
in the universal interaction and nonuniversal interactions of ${\it O}(\kf^4)$ is:
\beqa
{E}_{FL}/N \ = \ {\cal E}_{FL}/\rho \ &=&\   \varepsilon_{FG} \Big\lbrack 1\ +\ (g-1) \alpha \ +\ (g-1)\alpha^2\left(\frac{3}{4}-\log 2+ \frac{\pi}{8} \sigma_2 \kf^2 \right) \nonumber \\[4pt]
&+& \: (g-1) \alpha^3  \big\lbrack  0.16079 \ + \ (g-3)\left(\log 4 -1 \right)\big\rbrack  \nonumber \\[4pt]
 &+&\: (g+1) \frac{1}{\pi} \left(\sigma_p \kf^2\right) \ +\     (g-2)(g-1) \frac{1}{24\pi^2} \left(M D_0 \kf^2\right) 
\Big\rbrack \, .
   \label{energyperpartFL}
   \eeqa
where $\alpha \equiv \alpha(\kf)$.
Omitting nonuniversal effects, it is convenient to use the RG
evolution of $\alpha$, via Eq.~(\ref{eq:rg1sol}), to express the energy in
terms of the arbitrary scale $\lambda\kf$, where $\lambda$ is an
arbitrary real number\footnote{Note that the expression for the energy density with arbitrary $\lambda$ is precisely the expression that would
have been obtained if the scale $\mu$ had been kept arbitrary throughout the perturbative calculation.}. One obtains
\beqa
&&\!\!E_{FL}/N \ =\  \varepsilon_{FG} \Big\lbrack 1\ +\ (g-1) \alpha(\lambda\kf) \ +\ (g-1)\alpha(\lambda\kf)^2\left(\frac{3}{4}-\log 2\lambda \right) \nonumber \\
&&\!\!\!\!\!\!\!\!\!\!\!\!  + \: (g-1) \alpha(\lambda\kf)^3  \big\lbrack  0.16079  +  (g-3)\left(\log 4 -1 \right)-\left(\frac{3}{2}-\log 4\lambda\right)\log \lambda \big\rbrack + {\it O}(\alpha(\lambda\kf)^4 ) \Big\rbrack \ ,
   \label{energyperpartFLb}
   \eeqa
with
\begin{figure}[!h]
\centering
\includegraphics[width = 0.91\textwidth]{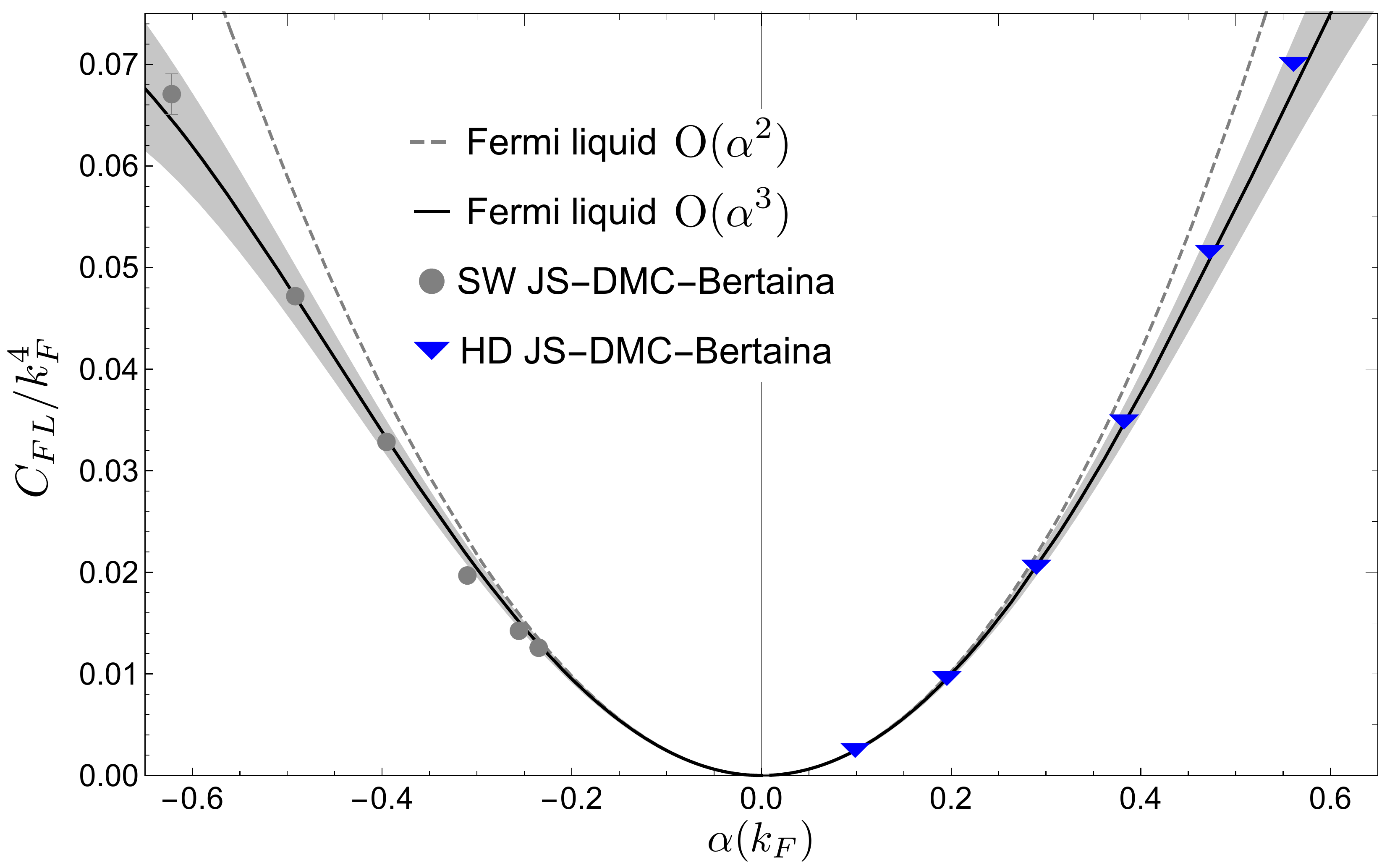}
\caption{The contact density, $C_{FL}$ in units of $\kf^4$, versus coupling
strength. The grey dashed curve is derived from the ${\it{O}}(\alpha^2)$ Fermi liquid energy and the solid black curve is from ${\it{O}}(\alpha^3)$. The gray band corresponds to varying the RG scale in the ${\it{O}}(\alpha^3)$ Fermi liquid energy and is a measure of the uncertainty associated with truncating the perturbative expansion, as discussed in the text. 
Note that the MC data have the gap contribution removed.}
\label{fig:contactnogap}
\end{figure}  
\begin{eqnarray}
\alpha(\lambda\kf) &=& -\left(\log{\left(\lambda \kf a_2 \right)}\right)^{-1} \ .
\label{eq:alphaarb}
\end{eqnarray}
   Note that
\beqa
\lambda \frac{d}{d \lambda }E_{FL}/N &=&  {\it O}\left(\alpha(\lambda\kf )^4\right) \ ,
   \label{energyperpartFL2rg}
\eeqa
and therefore any choice of $\lambda$ leads to the same physical prediction at the order in $\alpha$
computed. However, it is convenient to choose $\lambda$ to be consistent with the relevant physical scales and optimize perturbation theory. Below, the variation in $\lambda$ will be used to estimate the uncertainty due to neglecting higher orders in the perturbative expansion.

The contact~\cite{TAN20082971,TAN20082952,PhysRevA.86.013626} is an
observable of short-range interacting gases relating the derivative of
the energy with respect to the coupling constant to various static and
thermodynamic properties, such as the large momentum tail of the
momentum distribution or the high-frequency tail of relevant spectral
functions. The theoretical and experimental determination of the
contact has thus become a stringent test of internal consistency. Here
the s-wave\footnote{Contacts for the p-wave interaction and for the
  effective range may also be
  defined~\cite{PhysRevLett.115.135304,PhysRevA.95.023603,He_2021,He_2019,Zhang_2017,He_2016}.} contact density
is defined as
\begin{eqnarray}
         C &=& 2\pi M \frac{d{\cal E}}{d\ln \kf a_2} \ =\ 2\pi M \frac{d(\rho E/N)}{d\ln \kf a_2} \ .
\end{eqnarray}
In order to compare the prediction from Fermi liquid theory with MC simulation it is convenient to define the contact with the gap subtracted as $C_{FL}$.
Keeping universal interactions and assuming $g=2$ (two-component fermions) gives
\begin{equation}
  C_{FL}/\kf^4 = \frac{1}{4} \alpha^2\Big\lbrack 1 + \left(\frac{3}{2}-\log 4\right )\alpha + 3  \big\lbrack  0.16079 -  \left( \log 4 -1 \right)\big\rbrack \alpha^2
  + {\it O}(\alpha^3 ) \Big\rbrack  \ .
    \label{fig:Cont}
\end{equation}
This is plotted in Fig.~\ref{fig:contactnogap} where the gray shaded band corresponds to varying the RG scale by $10\%$ around the Fermi surface,
i.e. $\lambda=1\pm 0.05$. 
The comparison with MC simulations is discussed in Sec.~\ref{sec:MCS}.

\section{Ladders and rings to all orders}
\label{sec:LaR}
\begin{figure}[!h]
\centering
\includegraphics[width = 0.92\textwidth]{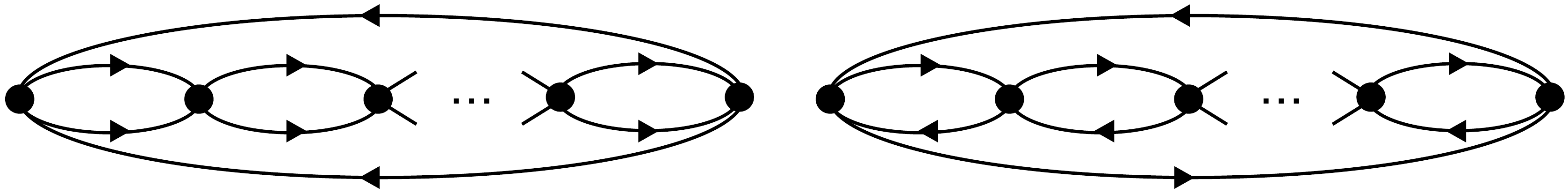}
\caption{Schematic representation of ladder diagrams to all orders (left) and ring diagrams to all orders (right). The black circle corresponds to an insertion of the
  $C_0$ operator.}
    \label{fig:Eladdring}
\end{figure}   

\noindent There have been many efforts to treat classes of diagrams to
all orders in the interaction strength $\alpha$, particularly in the
case of three dimensions~\cite{fetter1971quantum}, where such
resummations provide some insight regarding the energy density of the
Fermi gas at
unitarity~\cite{Steele:2000qt,Schafer:2005kg,Schafer:2006yf,Kaiser:2011cg,Kaiser:2012sr,Kaiser:2013bua}.

\begin{figure}[!h]
\centering
\includegraphics[width = 0.82\textwidth]{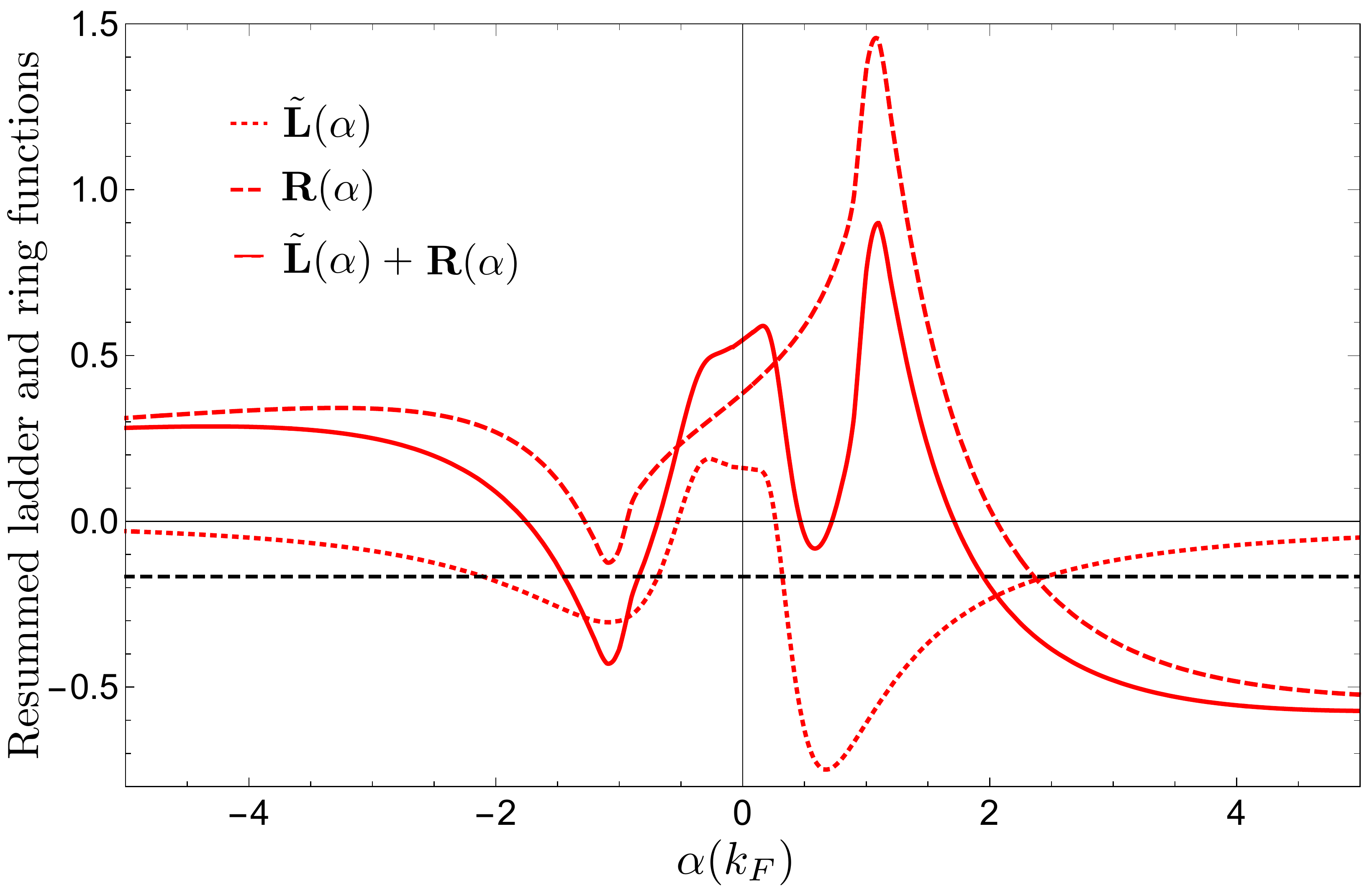}
\caption{Resummed ladder and ring functions, ${\tilde{\bf L}}(\alpha)$ and ${\bf R}(\alpha)$, respectively, and their sum, versus coupling strength. The dashed black line is the asymptotic value of the ring function, $-1/6$.}
    \label{fig:ringR}
\end{figure}

The complete ladder and ring diagram resummations have been computed for the Fermi gas in two
dimensions in Ref.~\cite{Kaiser:2013bua} (Appendix A) and in
Ref.~\cite{Kaiser:2014laa}, respectively, and the results will be
reproduced here for the purpose of comparison with MC simulations.  Fig.~\ref{fig:Eladdring} provides a schematic
illustration of the ladder and ring diagrams. The resummed energy density is
\beqa
E_{FL}/N &=&  \varepsilon_{FG} \Big\lbrack 1\ +\ {\bf L}(\alpha) \ +\ \alpha^3 {\bf R}(\alpha) \Big\rbrack \nonumber \\
& =& \varepsilon_{FG} \Big\lbrack 1\ +\ \alpha \ +\ \alpha^2\left(\frac{3}{4}-\log 2 \right) \ +\ \alpha^3\left({\bf R}(\alpha)+{\tilde{\bf L}}(\alpha)\right) \Big\rbrack \, ,
   \label{energyperpartLandRs}
   \eeqa
   where~\cite{Kaiser:2013bua}
\beqa
{\bf L}(\alpha) &=&  - \frac{32}{\pi} \int\displaylimits_{0}^{1} \! {d{s}} \, s \int\displaylimits_{0}^{\sqrt{1-s^2}} \! dt \, t\,\arctan{\frac{J(s,t)}{H(s,t)-\alpha^{-1}}}   \ ,
   \label{LadderSumenergyperpart}
   \eeqa
${\tilde{\bf L}}(\alpha)$ is defined by Eq.~(\ref{energyperpartLandRs}), and 
\beqa
{\bf R}(\alpha) &=&  -\frac{16}{\alpha^3 \pi} \int\displaylimits_{0}^{\infty}  \!dx\, x^2 \int\displaylimits_{\bar{\nu}_{\rm min}}^{x+1} \! 
d\bar{\nu}\, \bigg\{\alpha I(\bar{\nu},x)\big[\alpha R(\bar{\nu},x)+1\big] \nonumber \\ && + \: \arctan
{\alpha I(\bar{\nu},x)\over \alpha R(\bar{\nu},x)+2} +3 \arctan{\alpha I(\bar{\nu},x) \over \alpha 
R(\bar{\nu},x)-2} \bigg\} \ ,
   \label{RingSumenergyperpart}
   \eeqa
   where $\bar{\nu}_{\rm min} = {\rm max}(0,x-1)$, and $I(\bar{\nu},x)$ and $R(\bar{\nu},x)$ are defined above~\cite{Kaiser:2014laa}. This ring function satisfies the asymptotic
   conditions ${\bf R}(\pm\infty)=-1/6$, and both the ladder and ring functions and their sum are plotted in Fig.~\ref{fig:ringR}. Note that the resummed energy density agrees with the perturbative expansion up to ${\it O}(\alpha^3)$. It is therefore somewhat indicative of the
   uncertainty associated with the truncation of the perturbative expansion, as will be seen below. Beyond that, its
   implications, while interesting, are evidently academic and aspirational.

\section{Comparison with Monte Carlo simulations}
\label{sec:MCS}
\begin{figure}[!h]
\centering \includegraphics[width = 0.934\textwidth]{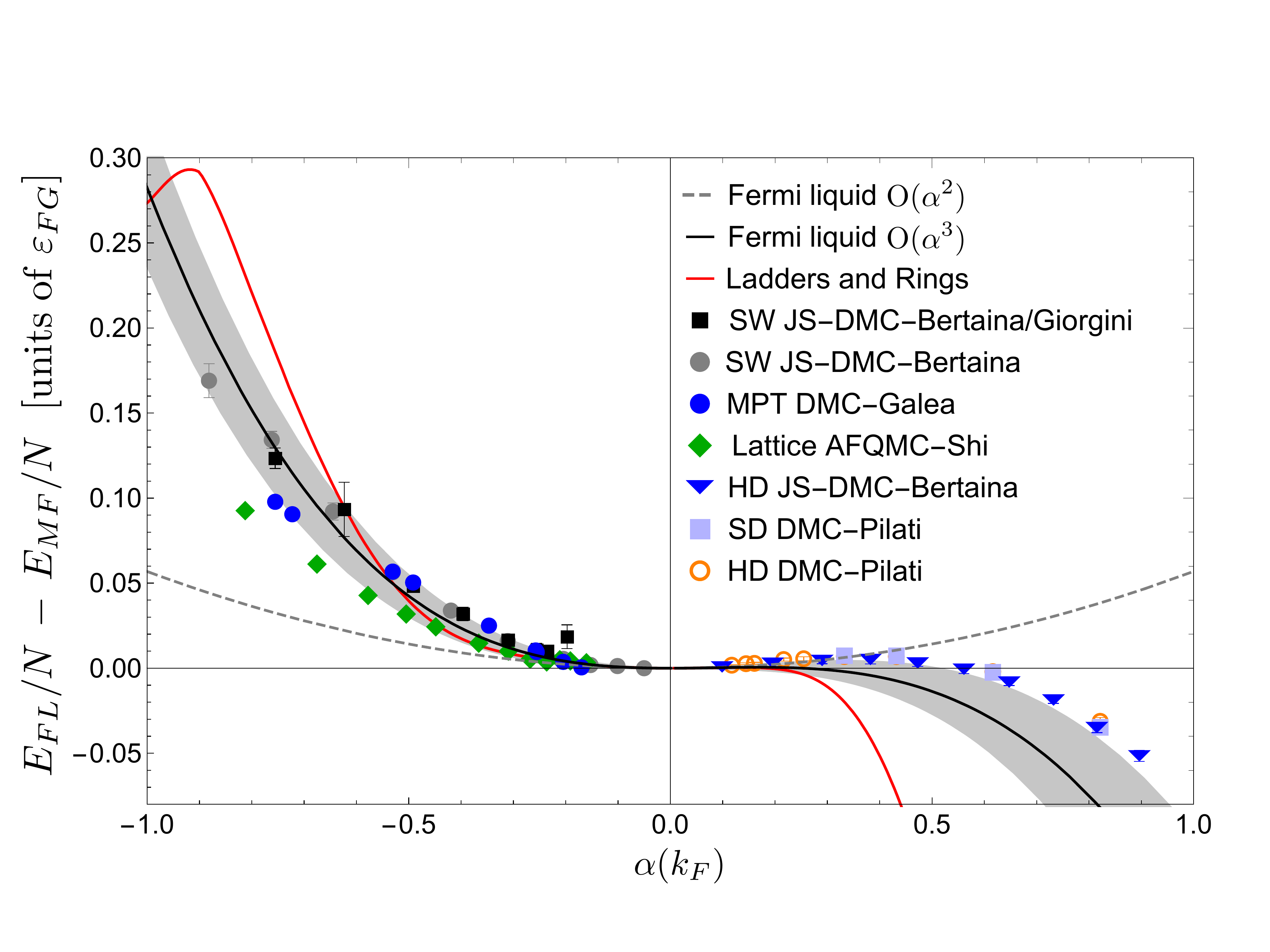}
\caption{Energy per particle with mean-field piece subtracted versus
  coupling strength. The grey dashed curve is NLO and the solid black
  curve is NNLO. The gray band corresponds to varying the
  RG scale at NNLO and is a measure of the uncertainty
  associated with truncating the perturbative expansion, as discussed
  in the text. The red curve is the complete ladder and ring
  resummation. The MC data are as described in the text.}
    \label{fig:epp1}
\end{figure}

\noindent Numerical simulations of the energy density of the
two-dimensional Fermi gas in the weakly-repulsive regime, and from the weakly-attractive BCS regime to the
strongly-coupled BEC regime, using MC techniques, have been
carried out in
Refs.~\cite{PhysRevLett.106.110403,BertainaS,PhysRevA.92.033603,Galea:2015vdy,Galea:2017jhe,Rammelmuller:2015ybu,PhysRevA.96.061601,PhysRevA.103.063314}.
The original study by Bertaina and
Giorgini~\cite{PhysRevLett.106.110403} used fixed-node diffusion Monte
Carlo (DMC). This study was then augmented by
Bertaina~\cite{BertainaS}, who increased the statistics in the
attractive branch, and also considered the repulsive branch.  Dramatic
improvements were then carried out by Shi {\it et
  al}~\cite{PhysRevA.92.033603}, who used auxiliary-field diffusion Monte Carlo (AFDMC), and achieved a more accurate nodal surface (guiding
wavefunction).  The results of Shi were then largely confirmed by
Galea {\it et al}~\cite{Galea:2015vdy}, who used fixed-node DMC with a refined nodal surface.

\begin{figure}[!h]
 \centering \includegraphics[width = 0.91\textwidth]{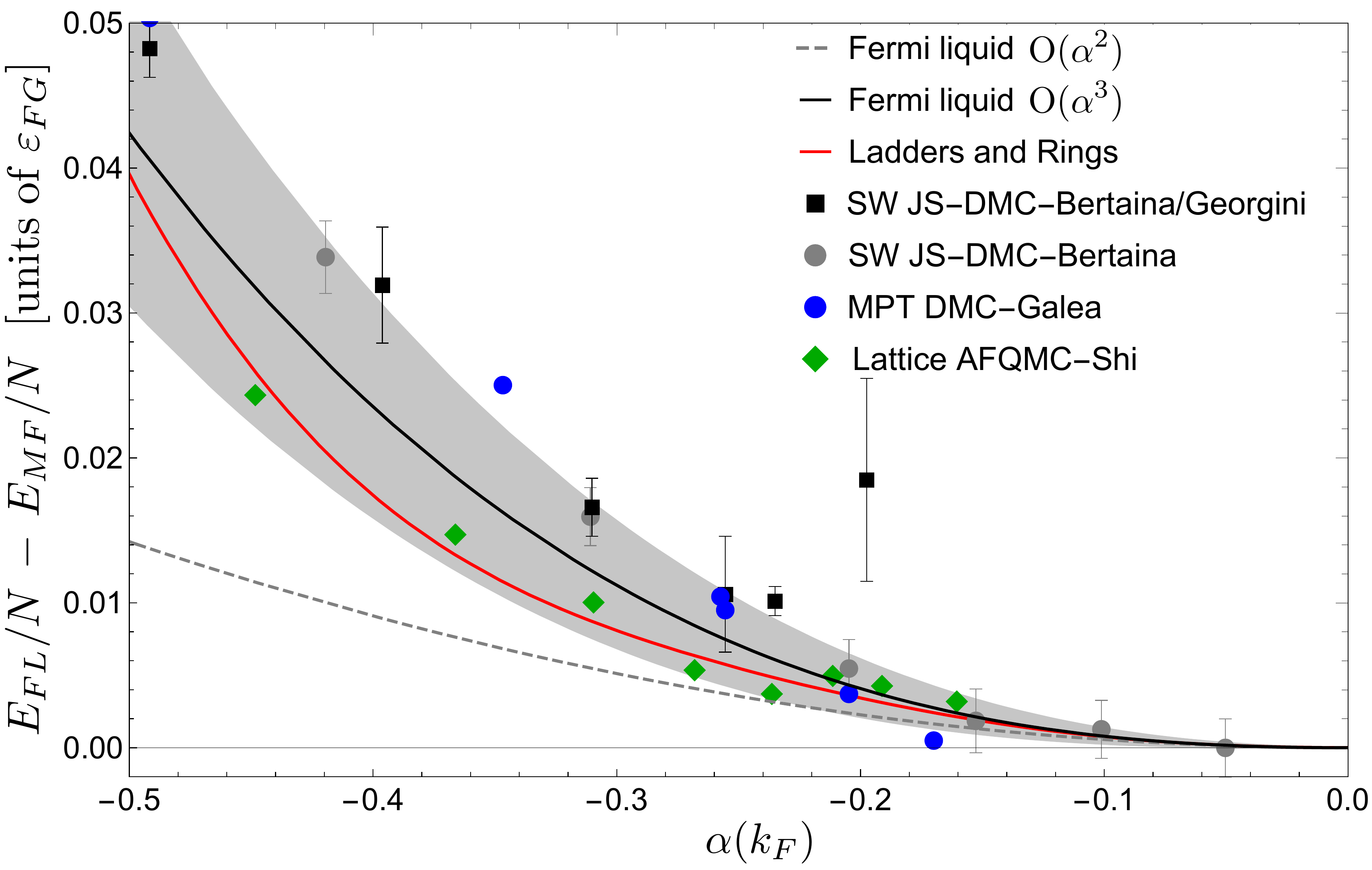}
 \caption{Energy per particle with mean-field contribution subtracted versus coupling strength. Magnification of Fig.~\protect\ref{fig:epp1} near the origin of the
   attractive branch.}
    \label{fig:epp1ins}
\end{figure}

With $g=2$ and omitting for now range corrections and other nonuniversal effects, which are mostly negligible in the simulations, one has from Eq.~(\ref{energyperpartFL}) the prediction
\beqa
E_{FL}/N &=&  \varepsilon_{FG} \Big\lbrack 1\ +\  \alpha \ +\ \alpha^2\left(0.05685 \right) \ -\  \alpha^3  \left(0.22550\right) \ +\ {\it O}(\alpha^4 ) \Big\rbrack \ .
   \label{energyperpartFL2}
   \eeqa
It is useful to define the mean-field contribution as $E_{MF}/N =
\varepsilon_{FG} \lbrack 1\ +\ \alpha \rbrack$.  Fig.~\ref{fig:epp1}
plots the energy-per-particle with the mean-field contribution
subtracted and shows that including the $\it{O}(\alpha^3)$ contribution does indeed restore the agreement between theory and MC simulation on the attractive side. Fig.~\ref{fig:epp1ins} magnifies this comparison for small negative $\alpha$. Note that all simulation data shown in the figures in the
attractive regime have the NLO gap energy subtracted, as is necessary to
compare with the Fermi liquid predictions, as discussed above and
indicated in Eq.~(\ref{Smatmobius1a}). In most MC simulation papers,
the energy density is expressed as a function of $\log{\left(\kf
  a_{2D}\right)}=\log{\left(\kf a_{2}\right)} -\gamma+\log 2$, and
therefore one must either translate between the two scattering length
conventions or use the freedom in changing the RG scale to achieve the
same effect.  The gray shaded band again corresponds to varying the
RG scale by $\lambda=1\pm 0.05$. The red curve is the complete ladder and
ring diagram sum.

Galea {\it et al} obtain smaller energies than Bertaina and Giorgini as the coupling increases,
a reflection of the more accurate nodal surface. Because fixed-node DMC is a
variational method, it is expected that the lower energies provide a
more accurate calculation of the ground state.  The method used by
Shi, AFDMC, is free of the sign problem in the spin-balanced case, meaning that in
principle they do not have problems of accuracy stemming from a variational
nodal surface, which affect the DMC method used in both Galea's and
Bertaina's papers. However, it includes the mapping from a lattice to a continuous model which strictly holds only in the low-energy regime. Another systematic source of error in all MC
simulations is the correction for finite-size effects, which assumes
Fermi liquid theory, and an effective mass equal to the
non-interacting case, $M^*=M$. This effect is accounted for by Bertaina and
Giorgini with an increased uncertainty.  In Shi's paper, the results
presented account for the finite-size correction.  However, 
subtracting the finite-size correction brings in additional assumptions: in particular,
the difference in energy between the finite-size system and
the thermodynamic limit is used, as calculated with BCS theory, which is not
exact.  In summary, Shi's results are affected by finite-size
inaccuracies, which they correct, and the lattice to continuous mapping, while Galea's and Bertaina's results
are affected both by variational inaccuracy in the nodal surface and
finite-size effects, which are corrected by both Bertaina and
Galea\footnote{The finite-size correction function implemented
by Galea {\it et al} is taken from Refs.~\cite{Shithesis,Galeathesis}.}.

\begin{figure}[!h]
 \centering \includegraphics[width = 0.92\textwidth]{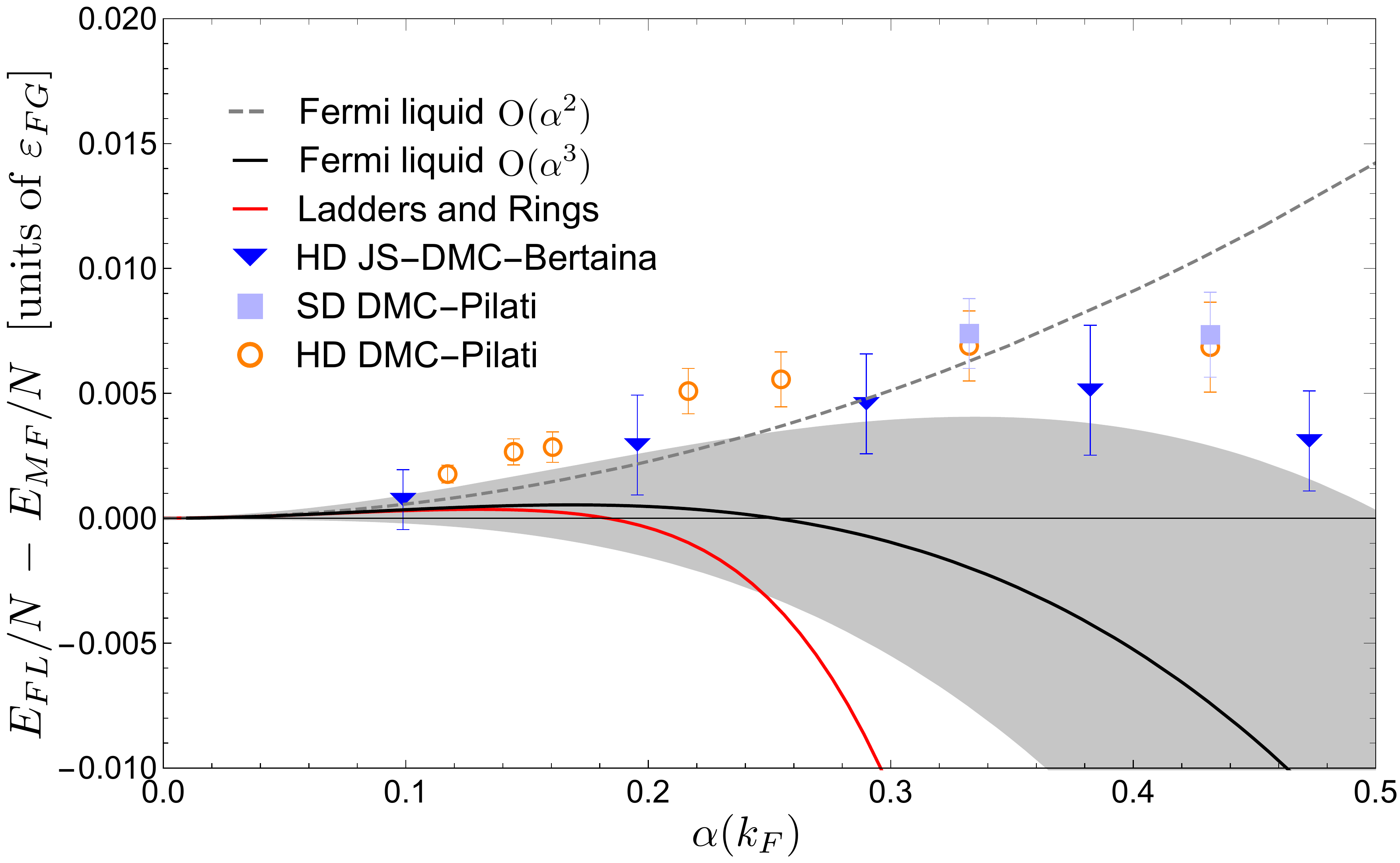}
 \caption{Energy per particle with mean-field contribution subtracted versus coupling strength. Magnification of Fig.~\protect\ref{fig:epp1} near the origin of the
   repulsive branch.}
    \label{fig:epp1insR}
\end{figure}

The repulsive two-dimensional Fermi gas has been studied with
fixed-node DMC in Refs.~\cite{BertainaS,PhysRevA.103.063314} for both
a hard-disk potential, where $a_{2D}$ is equal to the disk diameter
$R$, and a soft-disk potential where $a_{2D}=0.5R$. These results are
also reported in Fig.~\ref{fig:epp1} for $\alpha>0$, and are magnified in Fig.~\ref{fig:epp1insR} for small $\alpha$. 
Here, the fixed-node DMC  results are systematically slightly higher in energy than the uncertainty band of the EFT prediction.
This may be due to slower convergence of the perturbative expansion due to the beyond mean-field contribution alternating sign on the repulsive side.
Beyond the inaccuracies in the MC data associated with finite-size
effects (which are corrected as in the attractive case) and the
nodal surface of the trial wavefunction, nonuniversal effects may also be important in the repulsive regime. 
For example, in the case of a hard-disk potential, the s-wave effective range is comparable to the $a_{2D}$ parameter $\sqrt{|\sigma_2|}=a_{2D}/\sqrt{2\pi}$ \cite{Hammer:2010fw}.
In Figs.~\ref{fig:epp1} and \ref{fig:epp1insR}, for $\alpha(\kf)>0$,
the effective range contributions (computed for the hard-disk) are included in the Fermi liquid prediction at 
${\it O}(\alpha^3 )$. This amounts to a $5\%$ effect at $\alpha(\kf)\sim 1$. 
Note that the fixed-node DMC results for both the soft- and hard-disk potentials are consistent within error bars suggesting that effective range effects are not the major driver of discrepancy between fixed-node DMC and EFT. 
Both Figs.~\ref{fig:epp1ins} and \ref{fig:epp1insR} highlight that it would be worth pursuing new high precision MC simulations in the weakly-interacting regime, where both finite-size effects and effective range nonuniversality are accurately determined. This is particularly relevant in the repulsive case.  

The contact density predictions are compared with MC simulations of
the (short-range behavior of the) antiparallel pair distribution
function taken from Ref.~\cite{BertainaS} in
Fig.~\ref{fig:contactnogap}. The gap (molecular) contribution to the
contact density is subtracted from the data in the attractive regime.
Note that the contact density, unlike the energy per particle shown in Fig.~\ref{fig:epp1}, does not have the mean-field contribution subtracted.

\begin{figure}[!h]
 \centering  \includegraphics[width = 0.84\textwidth]{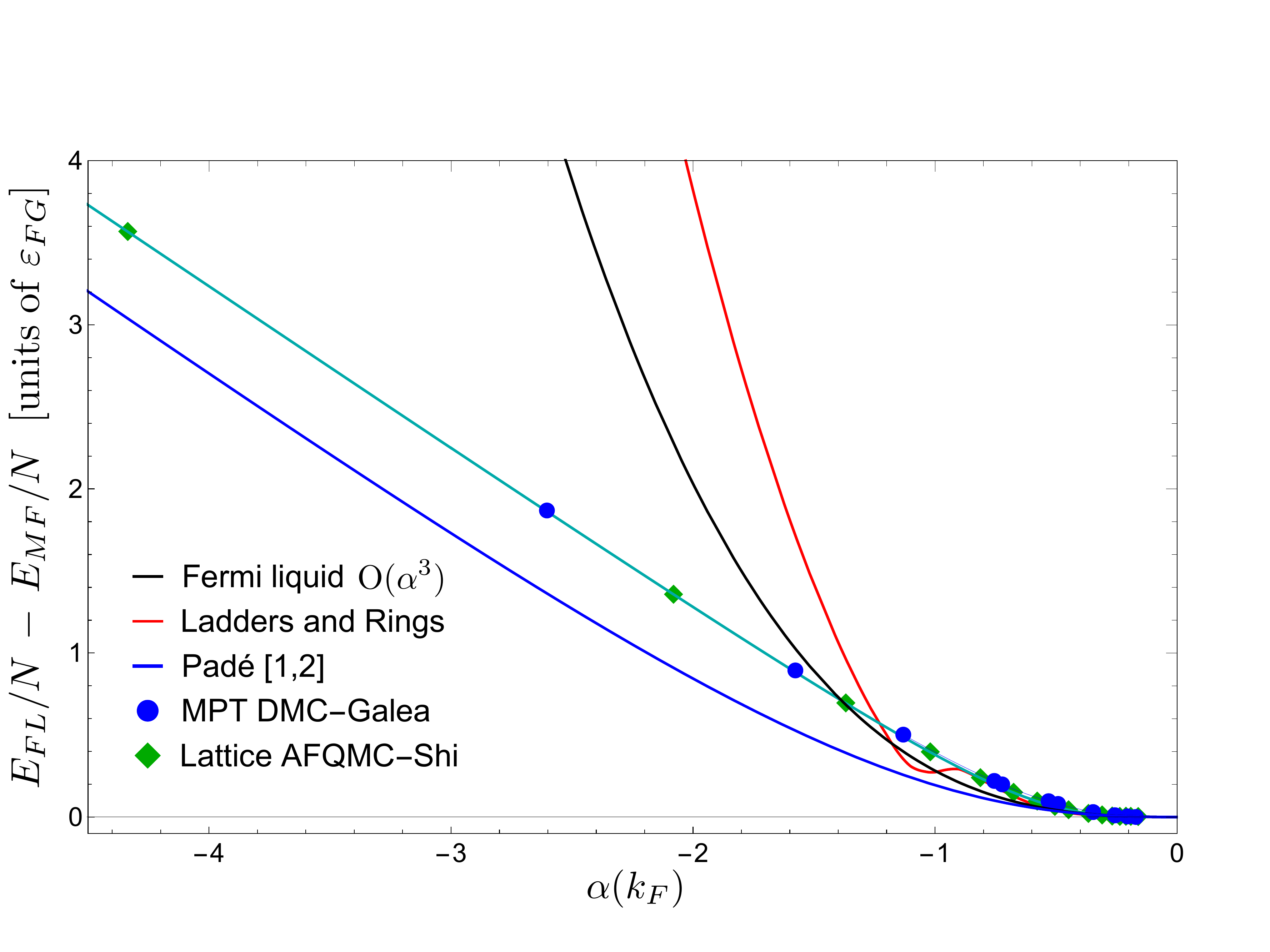}
 \caption{Energy per particle with mean-field piece subtracted versus
   coupling strength. MC data are included over all attractive coupling strengths
   from the BCS region to the BEC region. The data are as described in the text, with LO gap
   energy subtracted, and connected by the green curve for ease of
   viewing. The solid black curve is the
   NNLO Fermi liquid prediction.  The red curve is the complete ladder and ring resummation,
   and the blue curve is the [1,2] Pad\'e approximant, as discussed in
   the text.}
    \label{fig:padeepp1}
\end{figure}

Finally, Fig.~\ref{fig:padeepp1} shows the remarkably smooth and consistent MC data, 
here with the LO gap energy subtracted (see discussion at the end of Sec.~\ref{sec:gap}), out to
the strong coupling (BEC) region, where the fermions are expected to
form tightly bound pairs which in turn Bose condense. In the
three-dimensional case, resumming perturbation theory via Pad\'e
approximants and other methods appears to capture strong-coupling
trends fairly accurately~\cite{Wellenhofer:2020ylh}.  While the [2,1]
Pad\'e approximant formed from the results of this paper has a
low-lying singularity, the [1,2] Pad\'e approximant (shown in
Fig.~\ref{fig:padeepp1}) is much closer to the data than the
full ladder-ring sum. 

\section{Conclusion}
\label{sec:conc}

\noindent Quantum mechanics on a plane is remarkably rich and yet
dramatically distinct from its three dimensional counterpart, even in
the context of the ultra-cold, weak-coupling results that are considered in this
paper.  While challenging to realize experimentally, the weakly-coupled Fermi
gas in two dimensions is tractable analytically and can be simulated
to high accuracy using quantum MC techniques. In this paper, the universal
interaction has been computed to one order higher than known previously,
and the results have been shown to be in excellent agreement with MC simulations for attractive coupling.
In addition, various nonuniversal effects of interest have been computed, with
the hope that they will inspire specially-crafted MC simulations to test their
validity. The EFT methodology, with the choice of DR to tame
the singular nature of the interaction, proves to be a highly-efficient
means of systematically improving the description without having to specify the potential.

There are many straightforward generalizations of the results in this
paper. The most obvious extension is to compute one order higher in
the universal interaction. In three dimensions, the calculation of the
energy density has been taken to one higher
order~\cite{Wellenhofer:2018dwh,Wellenhofer:2021eis} in the
diagrammatic expansion, which is in correspondence with ${\it
  O}(\alpha^4 )$ effects in two dimensions. It is noteworthy
that at that order there are $\sim 30$ Feynman diagrams, most of which
are not of ring or ladder type. At this nominal fourth order,
resumming perturbation theory via Pad\'e approximants and other
methods appears to capture beyond perturbation-theory
strong-coupling trends accurately~\cite{Wellenhofer:2020ylh}.

Other interesting extensions of the work in this paper in the context
of the two-dimensional Fermi gas include the case of dilute Fermi
gases with population
imbalance~\cite{PhysRevA.103.063314,Foo:2019fre}, corrections to the
Fermi liquid quasiparticle
parameters~\cite{PhysRevB.45.10135,PhysRevB.45.12419}, which have been
computed to subleading orders in three
dimensions~\cite{Platter:2002yr}, and quantum corrections to the
energy density for the p-wave interactions (computed in three
dimensions in Ref.~\cite{Kaiser:2012sr}). In addition, the role of the
p-wave effective range in the many-body system is of current interest
both in two and three
dimensions~\cite{PhysRevLett.115.135304,PhysRevLett.123.070404,He_2021,He_2019,Zhang_2017,He_2016} and
may be profitably studied using the EFT methods of this paper.  Also
of interest are various corrections and extensions regarding the
pairing phenomena.  While the p-wave pairing gap was considered in the
original papers that addressed
superfluidity~\cite{PhysRevLett.62.981,PhysRevB.41.327}, the results
were somewhat mysterious due to the highly-singular nature of the
p-wave interaction.  It would be illuminating to address this problem
using EFT methods.

\section*{Acknowledgments}

\noindent We would like to thank E.~Berkowitz, David B.~Kaplan, Daniel R.~Phillips and Sanjay Reddy for helpful comments and discussions.
In addition, we are grateful to A.~Gezerlis and S.~Gandolfi for
making their data available, and for clarifying discussions regarding
their work.
G.B. acknowledges useful discussions with S. Pilati concerning inaccuracies in MC methods. This work was supported by the U.~S.~Department of Energy
grant {\bf DE-FG02-97ER-41014} (UW Nuclear Theory). R.F. was also partially supported by the U.~S.~Department of Energy
grant {\bf DE-SC0020970} (InQubator for Quantum Simulation).

\bibliographystyle{JHEP}
\bibliography{bibi}
\end{document}